\documentclass[12pt]{article}
\usepackage{vmargin}
\setpapersize[portrait]{A4}            
\setmargins{0.9in}{0.5in}{6.5in}{9.2in}
           {0.3in}{0.3in}{0.3in}{0.3in}
\usepackage{fancyhdr}
\usepackage{latexsym}
\usepackage{epsfig}
\usepackage{pstcol}
\usepackage{pst-text}
\usepackage{pst-node}
\usepackage{pstricks}
\usepackage{shadow,epsf,amsthm,amssymb,amsmath}
\usepackage{enumerate}
\usepackage{graphicx}
\usepackage{hvdashln}                           
\usepackage{setspace}                           

\newtheorem{definitionenv}{Definition}
\newtheorem{lemmaenv}[definitionenv]{Lemma}
\newtheorem{theoremenv}[definitionenv]{Theorem}
\newtheorem{corollaryenv}[definitionenv]{Corollary}
\newtheorem{propositionenv}[definitionenv]{Proposition}
\newtheorem{conjectureenv}[definitionenv]{Conjecture}
\newtheorem{exampleenv}{Example}
\newtheorem{app-lemmaenv}[section]{Lemma}

\newenvironment{definition}{\begin{definitionenv}\rm}{\end{definitionenv}}
\newenvironment{lemma}{\begin{lemmaenv}\rm}{\end{lemmaenv}}
\newenvironment{theorem}{\begin{theoremenv}\rm}{\end{theoremenv}}
\newenvironment{corollary}{\begin{corollaryenv}\rm}{\end{corollaryenv}}
\newenvironment{example}{\begin{exampleenv}\rm}{\end{exampleenv}}
\newenvironment{proposition}{\begin{propositionenv}\rm}{\end{propositionenv}}
\newenvironment{conjecture}{\begin{conjectureenv}\rm}{\end{conjectureenv}}
\newenvironment{app-lemma}{\begin{app-lemmaenv}\rm}{\end{app-lemmaenv}}

\newcommand{\bd}{\begin{definition}}
\newcommand{\ed}{\end{definition}}
\newcommand{\bl}{\begin{lemma}}
\newcommand{\el}{\end{lemma}}
\newcommand{\elp}{\hspace*{\fill} $\Box$
                 \end{lemma}}
\newcommand{\bt}{\begin{theorem}}
\newcommand{\et}{\end{theorem}}
\newcommand{\etp}{\hspace*{\fill} $\Box$
                 \end{theorem}}
\newcommand{\bc}{\begin{corollary}}
\newcommand{\ec}{\end{corollary}}
\newcommand{\ecp}{\hspace*{\fill} $\Box$
                 \end{corollary}}
\newcommand{\bcj}{\begin{conjecture}}
\newcommand{\ecj}{\end{conjecture}}
\newcommand{\be}{\begin{example}}
\newcommand{\ee}{\end{example}}
\newcommand{\eep}{\hspace*{\fill} $\Box$
                 \end{example}}
\newcommand{\bp}{\begin{proposition}}
\newcommand{\ep}{\end{proposition}}
\newcommand{\epp}{
                 \end{proposition}}

\newcommand{\ket}[1]{|#1\rangle}

\begin{document}

\title{
       A Construction of Quantum Stabilizer Codes Based on Syndrome Assignment by Classical Parity-Check Matrices
      }
\author{Ching-Yi Lai \ \ \mbox{and}\ \ Chung-Chin Lu
        \thanks{
        This work was supported by the
        National Science Council,
        Taiwan, under Contract NSC96-2628-E-007-006-MY3.
        The authors are with Department of Electrical Engineering,
        National Tsing Hua University,
        Hsinchu 30013, Taiwan. \
        E-mails:\quad cylai@abel.ee.nthu.edu.tw and
        cclu@ee.nthu.edu.tw.  Chung-Chin Lu is the person to
        correspond with.
        }
       }
\date{December 1, 2007}

\maketitle

\vspace{-1cm}
\begin{abstract}
In quantum coding theory, stabilizer codes are probably the most important class of quantum codes.
They are regarded as the quantum analogue of the classical linear codes and the properties of stabilizer codes have been carefully studied in the literature.
In this paper, a new but simple construction of stabilizer codes is proposed 
based on syndrome assignment by classical parity-check matrices.
This method reduces the construction of quantum stabilizer codes to the construction of classical parity-check matrices that satisfy a specific commutative condition.
The quantum stabilizer codes from this construction have a larger set of correctable 
error operators than expected.
Its (asymptotic) coding efficiency is comparable to that of CSS codes.
A class of quantum Reed-Muller codes is constructed, which have a larger set of correctable error operators than that of the quantum Reed-Muller codes developed previously in the literature.
Quantum stabilizer codes inspired by classical quadratic residue codes are also constructed and
some of which are optimal in terms of their coding parameters.
\end{abstract}

\begin{flushleft}
\end{flushleft}


\section{Introduction} \label{sec:introduction}
The theory of quantum error correction has been profoundly developed in the last decade
since the first quantum error-correcting code proposed by Shor \cite{Shor95}.
In \cite{Ste96a}, using a different approach from that of Shor, Steane gave
a new quantum error-correcting code and studied basic theory of quantum error correction.
Later, Steane gave more new quantum codes and discussed constructions of
quantum error-correcting codes in \cite{Ste96b}.
The complete quantum error correction condition was given in \cite{EM96,BDSW96,KL97}.
The optimal five qubit code was discovered in \cite{BDSW96,LMPZ96}.
After CSS code construction \cite{CS96,Ste96}, the study of quantum error-correcting codes
then turned to the study of classical self-orthogonal codes.

The idea of stabilizer codes was proposed in \cite{Got96,CRSS97} and
the properties of stabilizer codes were extensively addressed in
\cite{CRSS97,Got97}. CSS codes can then be viewed as one prominent class of stabilizer codes.
In \cite{Ste99}, Steane gave a further improvement, called an enlargement of CSS codes, which
produces several families of quantum codes with greater minimum distance.
In this paper we will propose a new but simple construction of quantum
stabilizer codes based on syndrome assignment by classical parity-check matrices and
develop several classes of quantum codes from this construction.

This paper is organized as follows.
In Section \ref{sec:stabilizer}, we begin in describing the basic properties of stabilizer codes
and end with a new formulation of CSS codes and their enlargement \cite{CS96,Ste96,Ste99}.
The method of construction of stabilizer codes based on syndrome assignment
by classical parity-check matrices is proposed
in Section \ref{sec:parity}, where we discuss the asymptotic coding efficiency of this method.
As an illustration, we develop a family of quantum stabilizer codes from classical Reed-Muller
codes in Section \ref{sec:rm}.
Several other quantum codes inspired by classical quadratic residue codes are
investigated in Section \ref{sec:cyclic}, where several optimal quantum codes are constructed.
A conclusion is discussed in the last section.

\section{Stabilizer Codes}
\label{sec:stabilizer}
\subsection{Stabilizer Groups and Stabilizer Codes}

Let $\cal H$ be the state space of a qubit.
The Pauli group $\mathcal{G}_n$, acting on the state space ${\cal H}^{\otimes {n}}$ of $n$ qubits,
plays an important role in the construction of $n$-qubit stabilizer codes.
An element in $\mathcal{G}_n$ is
expressed as $i^{c}M_1\otimes M_2\otimes \ldots \otimes M_n$,
where each $M_j$ is one of the Pauli operators $I$, $X$, $Y$, or $Z$ on $\cal H$,
$i=\sqrt{-1}$, and $c \in \{0,1,2,3\}$.
Let $\mathcal{K}=\{\pm I^{\otimes {n}}, \pm i I^{\otimes {n}}\}$, which
is a normal subgroup of $\mathcal{G}_n$ and
will be used in a later discussion.

For a $g\in \mathcal{G}_n$, the fixed subspace $\mathcal{V}(g)$ of $g$
is a subspace of ${\cal H}^{\otimes {n}}$ such that
$\ket{\psi} \in \mathcal{V}(g)$ if and only if $g\ket{\psi} =\ket{\psi}.$
A stabilizer group $\mathcal S$ that fixes a non-trivial subspace $\mathcal T$ of ${\cal H}^{\otimes {n}}$ is the set 
    \[
        \mathcal{S}=\left\{ g \in \mathcal{G}_n \left| g\ket{\psi }=\ket{\psi}, \ \ \forall \ket{\psi } \in \mathcal{T}\right.  \right\}.
    \]
A necessary condition is that $-I \notin \mathcal{S}$.
Since any $g,h\in \mathcal{G}_n$ have either $gh=hg$ or $gh=-hg$,
$\mathcal{S}$ must be an \emph{abelian} subgroup of $\mathcal{G}_n$.
And since any $g \in \mathcal{G}_n$ has either $g^2=I$ or $g^2= -I$,
we have $g^2=I \ \forall g \in \mathcal{S}.$
Therefore, $\mathcal{S} \cong(\mathbb{Z}_2)^r$ for some $r$, i.e., $\mathcal{S}=<g_1,g_2,\ldots ,g_r>$ with $r$ commutative independent generators.

An $[[n,k,d]]$ quantum stabilizer code $\mathcal{C}(\mathcal{S})$  is  a $2^k$-dimensional subspace of ${\cal H}^{\otimes {n}}$ fixed by a stabilizer group $\mathcal{S}$
with a set of $r=n-k$ independent generators.
The $d$ means the minimum distance of the quantum code $\mathcal{C(S)}$ and will be defined later.
The error-correction condition for a stabilizer code $\mathcal{C}(\mathcal{S})$
\cite{CRSS97,Got97,NC00} says that
$\{E_i \}$ is a collection of correctable error operators in $\mathcal{G}_n$ for $\mathcal{C}(\mathcal{S})$
if and only if
\begin{align}
    E_j^{\dag} E_k \notin N(\mathcal{S})\setminus \tilde{\mathcal{S}}\ \forall j,k, \label{eqn:qeccs}
\end{align}
where
    \[
        N(\mathcal{S})=\{ g\in \mathcal{G}_n \left| ghg^{\dag}\in \mathcal{S}
         \ \forall h \in \mathcal{S}\right. \}
    \]
is the normalizer group of $\mathcal S$ in $\mathcal{G}_n$, which in fact is the centralizer group
of $\mathcal S$ in $\mathcal{G}_n$, and
    \[
        \tilde{\mathcal{S}}=\mathcal{S}\mathcal{K}=\left\{ gh\left| \ g\in \mathcal{S},\ h\in \mathcal{K}\right. \right\}.
    \]
The weight of an element $i^{c} M_1\otimes M_2\otimes \ldots \otimes M_n$ in $\mathcal{G}_n $
is defined to be the number of $M_j$'s not equal to $I$.
Then the minimum distance $d$  of $\mathcal{C}(\mathcal{S})$,
motivated by the above error correction condition, is defined
to be the minimum weight of an element in $N(\mathcal{S})\setminus \tilde{\mathcal{S}}$.

\subsection{ Binary Codes Corresponding to Stabilizer Groups}
\label{sec:binaryStabilizer}
The Pauli group $\mathcal{G}_n$ is  closely related to the $2n$-dimensional binary vector space
$\mathbb{Z}_2^{2n}$. 
If $u, v$ are two binary $n$-tuples,  $(u,v)$ is meant to be a binary $2n$-tuples
and any element  $x \in \mathbb{Z}_2^{2n}$  can be written in the form $(u,v)$ with $u,v$ $n$-tuples.
We use $uv$ to denote the $n$-tuple of the bitwise AND of $u$ and $v$.
That is, $(uv)_{i}=u_i \cdot v_i$ where the subscript $i$ means the $i$-th bit of the binary $n$-tuple.
Then we define the generalized weight of an $2n$-tuple $x=(u,v)$ in $\mathbb{Z}_2^{2n}$, denoted  by $gw(x)$, as the Hamming weight of the bitwise OR of $u$ and $v$.
Thus \[gw(x)= w(u)+w(v)-w(uv),\]
where $w(u)$ means the Hamming weight of $u$, the number of nonzero components of $u$.

A $g=i^{c} M_1\otimes M_2\otimes \ldots \otimes M_n$ in $\mathcal{G}_n,$ can be expressed as $g=i^{c'} X_{\alpha}Z_{\beta}$, where $\alpha=(a_1, a_2, \ldots, a_n)$ and $\beta=(b_1, b_2, \ldots, b_n)$ are two binary $n$-tuples and $c,c' \in \{0,1,2,3\}$.
In this expression, if $M_j=I, X, Z, Y$, then $(a_j, b_j)=(0,0),(1,0),(0,1),(1,1)$, respectively.
And we have $c'\equiv c+l \ (\mbox{mod}\ 4)$ where $l$ is the number of $M_j$'s which are equal to $Y$
(note that $Y=iXZ$).
We define a group homomorphism $\tau : \ \mathcal{G}_n \mapsto \mathcal{G}_n/ \mathcal{K}  $ by
\[
    \tau(g)\triangleq g \mathcal{K} .
\]
If $g=i^c X_{\alpha}Z_{\beta}$, $\tau(g)=X_{\alpha}Z_{\beta} \ \mathcal{K}$.
Note that $X_{\alpha}Z_{\beta}\mathcal{K} = X_{\alpha'}Z_{\beta'}\mathcal{K} $
if and only if $\alpha =\alpha'$ and $\beta =\beta'$.
Also $\tau$ is an epimorphism.
Next we define a group isomorphism $\mu : \ \mathcal{G}_n/ \mathcal{K} \mapsto \mathbb{Z}^{2n}_2 $ by
\[
    \mu( X_{\alpha}Z_{\beta}  \mathcal{K})\triangleq (\alpha, \beta).
\]
Then we can define a homomorphism $\varphi : \ \mathcal{G}_n \mapsto \mathbb{Z}^{2n}_2 $ by $\varphi = \mu \circ \tau$, i.e.,
\[
    \varphi(i^c X_{\alpha}Z_{\beta})\triangleq\mu(\tau(i^c X_{\alpha}Z_{\beta}))=\mu( X_{\alpha}Z_{\beta}\mathcal{K})=(\alpha, \beta).
\]
It is clear that $\varphi$ is an epimorphism with kernel $\mathcal{K}$ and
$\{g_1,g_2,\ldots,g_r\}$ is a set of independent generators
if and only if $\varphi(g_1),\varphi(g_2),\ldots,\varphi(g_r)$ are linearly independent
$2n$-tuples in $\mathbb{Z}^{2n}_2$.

Suppose $\{ g_1, \ldots, g_r\}$ is a set of independent generators of a stabilizer group $\mathcal{S}$.
We define a \emph{check matrix} $H$ of $\mathcal{S}$
by making $\varphi(g_i) $ 
as its $i$-th row vector.
Then $H$ is an $r\times 2n$ binary matrix.
For convenience, we may denote $H$ by
\[H= \left[\begin{array}{c@{\vdashline}c} H_X  & H_Z  \end{array}\right],\]
where $H_X, H_Z$ are two $r\times n$ binary matrices.
Note that each of the $2^r$ sets $\{\pm g_1, \ldots, \pm g_r\}$ can be used as a set of independent generators of a distinct stabilizer group.
But all of them have the same check matrix $H$.

One example of stabilizer codes is a five-qubit code.
A set of generators of a $[[5,1,3]]$ code can be
         \begin{align*}
         g_1 &=\ XZZXI,\\
         g_2 &=\ IXZZX,\\
         g_3 &=\ XIXZZ,\\
         g_4 &=\ ZXIXZ.
        \end{align*}
The corresponding check matrix is
\begin{align}
    \begin{bmatrix}
    10010 & 01100 \\
    01001 & 00110 \\
    10100 & 00011 \\
    01010 & 10001 \\
    \end{bmatrix}.
    \label{eqn:5qubitcode}
\end{align}

Since a stabilizer group $\mathcal{S}$ is an abelian group, we have $ gh=hg \ \forall g, h\in \mathcal{S} $,
which has a corresponding property in $\varphi(\mathcal{S})$, induced by the homomorphism $\varphi$, that
    \[
        \varphi(g)\Lambda_{2n}\varphi(h)=0, \ \forall g, h\in \mathcal{S},
    \]
where $\Lambda_{2n} =\begin{bmatrix} O_{n\times n} & I_{n\times n}\\I_{n\times n}& O_{n\times n}\end{bmatrix}$.
Thus a check matrix $H$ of a stabilizer group $\mathcal{S}$ has to satisfy the
following \emph{commutative condition},
\begin{align}
    H \Lambda_{2n} H^T = H_X H_Z^T+H_ZH_X^T=O_{r\times r}, \label{eqn:commutative}
\end{align}
where $O_{i\times j}$ is the $i\times j$ zero matrix.
We will omit the subscripts of  $\Lambda $ and $O$ in the following discussion.
We define that  an $r\times 2n$ binary matrix $H=\left[ H_X \right. \left|H_Z\right]$ is \emph{commutative} if it satisfies the commutative condition.
From (\ref{eqn:commutative}), an $r\times 2n$ binary matrix $H=\left[ H_X \right. \left|H_Z\right]$ is \emph{commutative} if and only if $H_X H_Z^T$ is a symmetric $r\times r$ matrix.

The check matrix is a convenient tool for the encoding and decoding of
stabilizer codes \cite{Got97}.
In addition, the check matrix is able to
facilitate the construction of stabilizer groups from known
classical binary codes, as will be demonstrated in the next section.
Before that, we will illustrate an application of the check matrix
of a stabilizer group to determine the minimum distance of the
corresponding stabilizer code as follows.


Let $\bar{\mathcal S}= \varphi(\mathcal{S})$,
which is a subspace of $\mathbb{Z}_2^{2n}$.
Then $\bar{\mathcal S}$ can be viewed as a classical binary code with generator matrix $H$.
As in \cite{CRSS97,CRSS98}, we define a symplectic inner product $*$ on $\mathbb{Z}_2^{2n}$ by
\begin{align*}\label{eqn:sym inner product}
    (u_1,v_1)* (u_2,v_2)\triangleq u_1\cdot v_2+ v_1 \cdot u_2.
\end{align*}
Thus, two elements $g,h$ in $\mathcal{G}_n$ is commutative if and only if the symplectic inner product
$\varphi(g) * \varphi(h)$  of $\varphi(g)$ and $\varphi(h)$ is zero.
Let $\bar{\mathcal S}^{\bot *}$ be the dual code of $\bar{\mathcal S}$ with respect to the symplectic inner product,
    \[
        \bar{\mathcal S}^{\bot *} =\{ (u,v)\in  \mathbb{Z}_2^{2n} \left| (u,v)*(\alpha,\beta)=0,
        \ \forall (\alpha,\beta)\in \bar{\mathcal S}\right.\}.
    \]
It is clear that $\bar{\mathcal S}$ is a self-orthogonal code
with respect to this symplectic inner product, i.e.,
$\bar{\mathcal{S}} \subset \bar{\mathcal{S}}^{\bot *}$.
An $(2n-r)\times 2n$ binary matrix
    \[
        G= \left[\begin{array}{c@{\vdashline}c} G_X  & G_Z  \end{array}\right]
    \]
of rank $2n-r$, where $G_X$ and $G_Z$ are $(2n-r)\times n$ binary matrices,
is a generator matrix of $\bar{S}^{\bot *}$ if and only if
\begin{align} \label{eqn:commutative2}
    H\Lambda G^T=H_X G_Z^T+H_Z G_X^T=O.
\end{align}
It can be seen that $\bar{\mathcal S}^{\bot *}=\varphi(N(\mathcal{S}))$ and
the minimum distance of $\mathcal{C}(\mathcal{S})$ is just the minimum
generalized weight of a nonzero codeword in $\bar{\mathcal{S}}^{\bot *}\setminus \bar{\mathcal{S}}$.
This helps decide the minimum distance of a stabilizer code, as illustrated
in the construction of CSS codes
\cite{CS96,Ste96}
and their enlargement
\cite{Ste99}
as follows.

To construct an $[[n,k,d]]$ CSS code $\mathcal{C}(\mathcal{S})$,
we  choose a classical $[n,k_1]$ binary code $\mathcal{C}_1$ and
an $[n,k_2]$ subcode $\mathcal{C}_2$ of $\mathcal{C}_1$
such that both the code $\mathcal{C}_1$ and the classical dual code of $\mathcal{C}_2$
have minimum distance $\geqslant d$.
Then a check matrix of a stabilizer group $\mathcal{S}$ is established as
    \[
    H= \left[
        \begin{array}{c@{\vdashline}c}
            G_2 & O \\
            O & H_{1}
        \end{array}
        \right],
    \]
of rank $n-k_1+k_2$,
where $G_2$ is a generator matrix of $\mathcal{C}_2$ and $H_1$ is a parity-check matrix of $\mathcal{C}_1$.
Let $G_1=\left[ \begin{array}{c} G_2 \\ G_3 \end{array}\right]$, a generator matrix  of $\mathcal{C}_1$,
and  $H_2=\left[ \begin{array}{c} H_1 \\ H_3 \end{array}\right]$, a parity-check matrix  of $\mathcal{C}_2$.
Then a generator matrix of the symplectic dual code
$\bar{S}^{\bot *}$ is
    \[
        G=\left[ \begin{array}{c|c} G_2 & O \\ G_3 & O \\ O & H_1 \\O & H_3 \end{array}\right]
        =\left[ \begin{array}{c|c} G_1 & O \\ O & H_2  \end{array}\right],
    \]
which is of rank $n+k_1-k_2$.
It can be verified that both (\ref{eqn:commutative})and (\ref{eqn:commutative2}) hold.
The minimum distance of the quantum code is
no less than the minimum generalized weight of the symplectic dual code
$\bar{S}^{\bot *}$ and is clearly $\geqslant d$ from the structure of its generator matrix $G$.
The dimension of the quantum code is $k=n-(n-k_1+k_2) = k_1-k_2$.

The enlargement of CSS codes in \cite{Ste99} is based on the CSS construction and
exchanges code dimension for error-correcting capability.
With additional stabilizer generators, this enlargement increases
the minimum distance of the code by half.
We  give an explicit description of the enlargement in terms of nonsingular matrices
and a proof that this enlargement is able to increase the minimum distance of the code by half
in the following theorem.
\bt
\label{thm:uuv}
    Let $\mathcal{C}_1$ be a classical $[n,k_1,d_1]$ binary code
    which contains its classical dual $\mathcal{C}^{\bot}_1$.
    Furthermore, let $\mathcal{C}_1$ be able to be enlarged to $\mathcal{C}_2=[n,k_2,d_2]$, where $k_2>k_1$.
    Suppose that $G_{1}$, $G_2=\left( \begin{array}{c}G_{1}\\G_{3}\end{array}\right)$
    are generator matrices of  $\mathcal{C}_1$, $\mathcal{C}_2$,respectively,
    and $H_2$, $ H_1=\left( \begin{array}{c}H_{2}\\H_{3}\end{array} \right)$ are parity-check matrices of  $\mathcal{C}_2$, $\mathcal{C}_1$, respectively.
    If there exists a $(k_2-k_1) \times (k_2-k_1)$ nonsingular binary matrix $P$
    such that $I+P$ is also nonsingular,
    by  taking
    \[
        H=\left( \begin{array}{c|c}H_{2} & O\\O & H_{2} \\ QH_3 & H_3 \end{array}\right),
    \]
    where $Q= \left(H_3 G_3^{T}\right) \left(P^{T}\right)^{-1} \left(H_3 G_3^{T}\right)^{-1}$
    is a $(k_2-k_1) \times (k_2-k_1)$ nonsingular matrix,  
    as a check matrix of a stabilizer group $\mathcal{S}$,
    an $[[n,k_2+k_1-n, d\geqslant\min\{d_1,\lceil \frac{3d_2}{2}\rceil\}]]$ quantum code $\mathcal{C}(\mathcal{S})$ can be constructed.
    A generator matrix of $\bar{S}^{\bot *}$ can be
    \[
        G= \left(\begin{array}{c@{\vdashline}c} G_1  & O \\ O& G_1 \\G_{3} & P G_{3}  \end{array}\right) .
    \]
\et
\begin{proof}
It is easy to see that the two matrices $H$ and $G$ are of full rank
and have rank $(2n-k_1-k_2)$ and rank $k_1+k_2$, respectively.
Since  $\mathcal{C}^{\bot}_1 \subset \mathcal{C}_1$, we have $H_1H_1^{T}=O$ and hence $H_2H_2^{T}=O$, $H_3H_3^{T}=O$, $H_2H_3^{T}=O$.
Thus
\begin{align*}
    H \Lambda H^T =& \left( \begin{array}{c}H_{2}\\ O\\ QH_{3}\end{array} \right)\left( \begin{array}{ccc}O^T &H_{2}^T &H_{3}^T\end{array} \right)
                            +\left( \begin{array}{c}O\\ H_{2}\\H_{3}\end{array} \right)\left( \begin{array}{ccc}H_{2}^T & O^T& H_{3}^TQ^T\end{array} \right)\\
                        =&  \left( \begin{array}{ccc}H_{2}O^T& H_{2}H_{2}^T&H_{2}H_{3}^T \\OO^T& OH_{2}^T& OH_{3}^T\\ QH_{3}O^T& QH_{3}H_{2}^T& QH_{3}H_{3}^T\end{array} \right)
                            +\left( \begin{array}{ccc}OH_{2}^T& OO^T&  OH_{3}^TQ^T\\H_{2}H_{2}^T& H_{2}O^T& H_{2}H_{3}^TQ^T\\H_{3}H_{2}^T&  H_{3}O^T& H_{3}H_{3}^TQ^T\end{array} \right)\\
                        =&  O+O = O,
\end{align*}
and Eq. (\ref{eqn:commutative}) holds.
Since $H_1G_1^{T}=O$ and $H_2G_2^{T}=O$, we have  $H_2G_1^{T}=O$, $H_3G_1^{T}=O$,
and $H_2G_3^{T}=O$. Thus
\begin{align*} 
    H\Lambda G^T=& \left( \begin{array}{c}H_{2}\\ O\\ QH_{3}\end{array} \right)\left( \begin{array}{ccc}O^T &G_{1}^T & G_{3}^TP^T\end{array} \right)
    +\left( \begin{array}{c}O\\ H_{2}\\H_{3}\end{array} \right)\left( \begin{array}{ccc}G_{1}^T & O^T& G_{3}^T\end{array} \right)\\
                        =&  \left( \begin{array}{ccc}H_{2}O^T& H_{2}G_{1}^T& H_{2}G_{3}^TP^T \\OO^T& OG_{1}^T& OG_{3}^TP^T\\ QH_{3}O^T& QH_{3}G_{1}^T& QH_{3}G_{3}^TP^T\end{array} \right)
                            +\left( \begin{array}{ccc}OG_{1}^T& OO^T&  OG_{3}^T\\H_{2}G_{1}^T& H_{2}O^T& H_{2}G_{3}^T\\H_{3}G_{1}^T&  H_{3}O^T& H_{3}G_{3}^T\end{array} \right)\\
                         =&  \left( \begin{array}{ccc}O& O& O \\O& O& O\\ O&O&   \left(H_3 G_3^{T}\right) \left(P^{T}\right)^{-1} \left(H_3 G_3^{T}\right)^{-1} H_{3}G_{3}^TP^T+H_{3}G_{3}^T\end{array} \right)\\
                        =&\ O,
\end{align*}
and Eq.~(\ref{eqn:commutative2}) holds.
The minimum distance $d$ of $\mathcal{C}(\mathcal{S})$ is determined as follows.
Consider a nonzero codeword
$(c_1,c_2)= (u_1|u_2|u_3)\left(\begin{array}{c@{\vdashline}c} G_1  & O \\ O& G_1 \\G_{3} & P G_{3}
   \end{array}\right)=(u_1G_1+u_3G_3,u_2G_1+u_3PG_3)$ in $\bar{\mathcal S}^{\bot *}$,
with two $k_1$-tuples $u_1,u_2$ and a $(k_2-k_1)$-tuple $u_3$, not all zero tuples.
It is clear that  $c_1,c_2\in \mathcal{C}_2$ and $w(c_1)\geqslant d_2, w(c_2)\geqslant d_2$.
If $c_1\neq c_2$, then $w(c_1+c_2)\geqslant d_2$ and we have
\begin{align*}
    gw(c_1,c_2)&= w(c_1)+w(c_1)-w(c_1c_2)\\ 
                &= \frac{1}{2}\left( w(c_1)+w(c_2)+ (w(c_1)+w(c_2)-2w(c_1c_2))\right)\\
                &=\frac{1}{2}\left( w(c_1)+w(c_2)+ w(c_1+c_2)\right)
                  \geqslant \frac{3d_2}{2}.
\end{align*}
Now if $c_1=c_2$, then
$u_1G_1+ u_3G_3 = u_2G_1+ u_3PG_3$ and then
$(u_1+u_2)G_1+ u_3G_3 = u_3PG_3$.
Since $G_1$ and $G_3$ have linearly independent rows, we must have
$u_1+u_2=0$ and $u_3G_3=u_3PG_3$.
However, since $I+P$ is nonsingular,  $u_3G_3 = u_3PG_3$ if and only if $u_3=0$.
Since $u_1,u_2,u_3$ are not all zero tuples, we have nonzero $u_1=u_2$ and
$c_1=u_1G_1\in \mathcal{C}_1,c_2=u_2G_1=u_1G_1=c_1$.
Thus $gw(c_1,c_2)=w(c_1)\geqslant d_1$.
In conclusion, the minimum distance of the quantum code is
\[d\geqslant \min\{d_1,\left\lceil \frac{3d_2}{2}\right\rceil\}.\]
\end{proof}

\section{A Simple Construction of Stabilizer Codes}
\label{sec:parity}

\subsection{Syndrome Assignment and Check Matrices}

Given a stabilizer group $\mathcal{S}$ together with a set of independent
generators $\{g_1,g_2,\ldots ,g_r\}$, the encoding-decoding techniques
of the corresponding stabilizer code is well studied in \cite{Got97}.
But one may ask the question: how to find a set of independent
generators $g_1,g_2,\ldots ,g_r$ such that the generated
stabilizer group will correspond to a  good quantum  code?

In \cite{Got96}, Gottesman  constructed a class of stabilizer
codes saturating the quantum Hamming bound which says that
    \begin{align*}
        2^k \sum_{i=0}^{t} 3^i {n \choose i} \leqslant 2^n.
    \end{align*}
These codes encode $k= n-j-2 $ in $n=2^j$ qubits and correct up to
$t=1$ error.
He assigned an ``error syndrome" to each correctable
error operator according to a certain rule so that error syndromes
form a check matrix of the stabilizer group.
This is a brilliant idea of giving a set of independent generators $g_1,g_2,\ldots ,g_r$.
We will basically follow this idea  to construct  $[[n,k,d]]$ quantum codes for the
general case of $t=\lfloor \frac{d-1}{2}\rfloor \geq 1$.

\subsubsection{Error Syndromes} \label{sec:architecture}
In classical coding theory,  correctable error patterns are a collection of  error patterns which have distinct error syndromes. 
The same concept holds in non-degenerate quantum codes and hence can be applied to construct non-degenerate quantum stabilizer codes.
In stabilizer codes, the idea of error syndrome comes from the commutativity.
The error-correction condition for stabilizer codes in (\ref{eqn:qeccs}) says that the multiplication of any two
correctable error operators in $\mathcal{G}_n$, each with weight no more than $t$, is not in $N(\mathcal{S})-\tilde{\mathcal{S}}$.
So each correctable error operator is assigned with a binary pattern indicating the commutative relation between the error operator and each one of the generators of a stabilizer code.
For our purposes, we will take a stricter error-correction condition that  multiplication of any two correctable error operators,
each with weight less than or equal to $t$,
must anti-commute with some element in $\mathcal{S}$.

As in \cite{Got96}, for any $g\in \mathcal{G}_n$, we define $f_g: \ \mathcal{G}_n \mapsto \mathbb{Z}_2$ by
    \[
    f_g(h)=\left\{%
            \begin{array}{ll}
                0, & \hbox{if \ [g,h]=0,} \\
                1, & \hbox{if \ \{g,h\}=0,} \\
            \end{array}%
            \right.
    \]
    where $[g,h]=gh-hg$ and $\{g,h\}=gh+hg$.
It can be verified that $f_g$ is a group homomorphism.
Then for a given set of independent generators $g_1$, $g_2$, $\ldots$, $g_r$  of a stabilizer group $\mathcal{S}$, we define $f_{\mathcal{S}}: \ \mathcal{G}_n \mapsto (\mathbb{Z}_2)^r$ by
\[
    f_{\mathcal{S}}(h)=\left(f_{g_1}(h),f_{g_2}(h),\ldots ,f_{g_r}(h)\right)^T.
\]
It is clear that
\begin{align}
    f_{\mathcal{S}}(h)=H\Lambda \varphi(h)^T, \label{eq:syndrome}
\end{align}
where $H$ is the check matrix of $\mathcal S$ corresponding to the (ordered)
generators $g_1, g_2, \ldots, g_r$.
Note that  $f_{\mathcal{S}}$ is a group homomorphism.
It is also clear that $f_{\mathcal{S}}(h)=(0,0,\ldots ,0)^T$ if and only if $h$ commutes with every element in $\mathcal{S}$, i.e., $h\in N({\mathcal{S}})$.
We call $f_{\mathcal{S}}(E)$  the \emph{error syndrome} of $E$ for each error operator $E$ in $\mathcal{G}_n$.
The error syndrome in this definition is equivalent to that defined in \cite{NC00} with $+1,-1$ in place of $0,1$, respectively.
For any two correctable error operators $E_1$ and $E_2$ in $\mathcal{G}_n$, each with weight less than or equal to $t$,
we attempt to construct a stabilizer group $S$ such that $f_{\mathcal{S}}(E_1E_2)$ is a nonzero vector.
Since $f_{\mathcal{S}}$ is a  group homomorphism,
        \[
            f_{\mathcal{S}}(E_1E_2) \neq 0 \Leftrightarrow f_{\mathcal{S}}(E_1) \neq f_{\mathcal{S}}(E_2).
        \]
As a result, we need to assign distinct error syndromes  to distinct correctable error operators.

If we can correct  error operators  $ X_1,\ldots, X_n, Y_1, \ldots, Y_n, Z_1, \ldots, Z_n$ and the multiplication of any no more than $t$ of them, we are able to correct all error operators up to weight $t$.
We call these $3n$ error operators to be basic correctable error operators.
Moreover, since the set $\{X_1,$ $X_2,$ $\ldots ,$ $X_n,$ $Z_1,$ $Z_2,$ $\ldots ,$ $Z_n\}$ generates all the error operators in $\mathcal{G}_n$ by multiplication up to a scalar factor,
it is desirable to determine the error syndromes of these $2n$ basic correctable error operators so that  the  error syndromes of all correctable error operators are distinct from each other.
For the case of $t=1$, it has been done in \cite{Got96}.
For the general case of $t\geqslant 2$, it becomes challenging to assign error syndromes to the $2n$ basic correctable error operators so that the error syndromes of correctable error operators of weight no more than $t$ are all distinct.

From (\ref{eq:syndrome}), we observe that the first $n$ columns and the last $n$ columns of the check matrix of a stabilizer group $\mathcal{S}$ corresponding to the (ordered) independent generators $g_1,g_2, \ldots, g_r$
are $f_{\mathcal{S}}(Z_1)$, $f_{\mathcal{S}}(Z_2)$, $\ldots $,$f_{\mathcal{S}}(Z_n)$ and $f_{\mathcal{S}}(X_1)$, $f_{\mathcal{S}}(X_2)$, $\ldots$, $f_{\mathcal{S}}(X_n)$, respectively.
Thus we can establish a check matrix of a target stabilizer group $\mathcal{S}$ by assigning $2n$ error syndromes as its columns
$f_{\mathcal{S}}(Z_1)$, $f_{\mathcal{S}}(Z_2)$, $\ldots $,$f_{\mathcal{S}}(Z_n)$, $f_{\mathcal{S}}(X_1)$, $f_{\mathcal{S}}(X_2)$, $\ldots$, $f_{\mathcal{S}}(X_n)$
and verifying this matrix to be commutative.
In this way, the method of {syndrome assignment} is just to define a {\em legal} check matrix.


\subsubsection{Syndrome Assignment by a Binary Parity-Check
Matrix} \label{sec:complexity}

Let $E = X_{\alpha }Z_{\beta }$ be an error operator of weight no more than $t^*$,
where $\alpha =( a_1, \ldots, a_n )$, $\beta =(b_1, \ldots, b_n)$
and $gw(\alpha, \beta)\leq t^* $.
Then for a target stabilizer group $\mathcal{S}$,
\[
    f_{\mathcal{S}}(E) = f_{\mathcal{S}}(X_{\alpha }Z_{\beta })= \sum_{i=1}^n a_i  f_{\mathcal{S}}(X_i) + \sum_{i=1}^n b_i  f_{\mathcal{S}}(Z_i),
\]
which is a linear combination of at most $2t^*$ terms.
To ensure that the error syndromes of two distinct error operators
$E_1=X_{\alpha_1 }Z_{\beta_1 }$ and $E_2=X_{\alpha_2 }Z_{\beta_2 }$ with
$\alpha_1 =( a_1, \ldots, a_n )$, $\beta_1=(b_1, \ldots, b_n)$, $gw(\alpha_1, \beta_1)\leq t^* $ and $\alpha_2 =( u_1, \ldots, u_n )$,$\beta_2 =(v_1, \ldots, v_n)$, $gw(\alpha_2, \beta_2)\leq t^* $ are distinct, we must have
    \begin{align}
        f_{\mathcal{S}}(E_1) \neq f_{\mathcal{S}}(E_2) &\Leftrightarrow f_{\mathcal{S}}(E_1)+f_{\mathcal{S}}(E_2) \neq 0  \nonumber  \\
        &\Leftrightarrow \sum_{i=1}^n (a_i+u_i)  f_{\mathcal{S}}(X_i) + \sum_{i=1}^n (b_i+v_i)  f_{\mathcal{S}}(Z_i)  \neq 0, \label{4t}
    \end{align}
which is a linear combination of at most $4t^*$ terms.
Thus a sufficient condition to guarantee the distinction among error syndromes of error operators of weight no more than $t^*$  is
 that any $4t^*$ elements in the set $\{f_{\mathcal{S}}(X_1),$ $f_{\mathcal{S}}(X_2),$ $\ldots ,$ $f_{\mathcal{S}}(X_n),$ $f_{\mathcal{S}}(Z_1),$ $f_{\mathcal{S}}(Z_2),$ $\ldots ,$ $f_{\mathcal{S}}(Z_n)\}$ must be linearly independent.
Surprisingly, this is just a property of a parity-check matrix of a classical linear block code with minimum distance $d'\geqslant 4t^*+1$,
where any $d'-1$ column vectors of the parity-check matrix must be linearly independent.

From the above discussion, we now know how to do syndrome assignment so that all correctable error operators will have distinct error syndromes from each other.
We first choose a classical $[2n,n+k,d']$ binary linear block code $\mathcal{C}'$ with $d'\geq 4t^*+1$ and $k>0$.
Let  $H'$ be a parity-check matrix of $\mathcal{C}'$ with dimension $(n-k) \times 2n$ .
Then the $2n$ column vectors of $H'$ will be assigned as $f_{\mathcal{S}}(X_i)$'s and $f_{\mathcal{S}}(Z_i)$'s.
There are $(n-k)$ independent generators of the target stabilizer group $\mathcal{S}$ and the corresponding target check matrix $H$ is just the permutation of the column vectors of $H'$,
i.e., $H=H'P$ for some permutation matrix $P$. Then the commutative condition becomes
    \begin{align}
        H'P \Lambda P^T H'^T =O. \label{eq:HP}
    \end{align}
If $G'$ is a generator matrix of $\mathcal{C}'$, then the symplectic dual
$\bar{\mathcal{S}}^{\bot*}$ of $\bar{\mathcal{S}}$ in $\mathbb{Z}_2^{2n}$ has a generator matrix
    \[
        G=G'P\Lambda
    \]
since
    \[
        H\Lambda G^T= H'P\Lambda \Lambda ^T P^T G'^T =H' G'^T =O.
    \]

If the target check matrix  $H$ satisfies (\ref{eq:HP}), then $H$
is indeed a check matrix of a stabilizer group which corresponds
to a quantum code that is at least $t^*$-error correcting. The dimension of the
quantum code is $n-(n-k)= k $. Thus the choice of $k>0$ ensures
that the corresponding quantum code is of dimension greater than
zero.

We conclude the above discussion  in the following theorem.
    \bt
        Given an $(n-k) \times  2n $ parity-check matrix $H'$ of a binary $[2n,n+k,d']$ linear block code $\mathcal{C}'$ with minimum distance $d'\geqslant 4t^*+1$,
        such that (\ref{eq:HP}) holds for a certain permutation $P$,
         an $[[n,k, d\geq 2 t^*+1 ]]$ stabilizer code with a check matrix $H=H'P$  can  be constructed.
         \label{thm:H}
    \et

A $t^*$-error-correcting quantum code of length $n$ has
    \[
        \sum_{i=0}^{t^*} {n\choose i}3^{i}
    \]
error syndromes.
One may expect that a quantum code by the above construction can correct more than just those error operators of weight$\leqslant t^*$ when $t^*$ is determined from $d'$ in Theorem $\ref{thm:H}$.
In fact, any error operator $E = X_{\alpha}Z_{\beta}$ with $w(\alpha)+w(\beta)\leqslant 2t^*$ has its own unique syndrome and then can be corrected.
For example,  error operators  $X_{\alpha}$ and $Z_{\alpha}$ with $w(\alpha)= 2t^*$ are correctable.

\begin{figure}
    \begin{center}
    \begin{pspicture}(0,0)(10,6)
        \pnode(0,0){p1}
        \pnode(0,5.5){p2}
        \pnode(8.5,0){p3}

%

        \rput(9,-0.25){\rnode{op1}{$m_{X,Z}$}}
        \rput(-0.25, 5.5){\rnode{op2}{$m_Y$}}
        \rput(-0.5, -0.5){\rnode{op1}{$0 $}}
        \rput(-0.5, 4){\rnode{op1}{$t^*+1 $}}
        \rput(4, -0.5){\rnode{op1}{$t^*+1 $}}
        \rput(-0.5, 5){\rnode{op1}{$t+1 $}}
        \rput(5, -0.5){\rnode{op1}{$t+1 $}}
        \rput(8, -0.5){\rnode{op2}{$2t^* $}}
        \rput(1, 1.5){\rnode{op3}{$m_Y+m_{X,Z}=t^*+1 $}}
        \rput(2, 4.25){\rnode{op3}{$m_Y+m_{X,Z}=t+1$}}
        \rput(7,1.5 ){\rnode{op3}{$2m_Y+m_{X,Z}=2t^*$}}
        \psline[fillstyle=vlines, linewidth=0](5,0)(8,0)(2,3)(5,0)
        \ncline{->}{p1}{p3}
       \pstextpath[c](0,-0.5){\psline[linestyle=dashed](0,4)(4,0)}{ }
        \pstextpath[r](0,0.4){\psline[linestyle=dashed](0,4)(8,0)}{ }
        \pstextpath[l](0,0.4){\psline[linestyle=dotted](0,5)(5,0)}{ }

        \ncline{->}{p1}{p2}

        \end{pspicture}
        \end{center}
  \caption{The $m_{X,Z}-m_Y$ region of additional correctable error operators. The $m_{X,Z}$-axis represents the number of components $M_j$'s of error operator $E$ equal to $X$ or $Z$ and the $m_Y$-axis represents the number of  components $M_j$'s equal to $Y$.}\label{fig:add}
\end{figure}
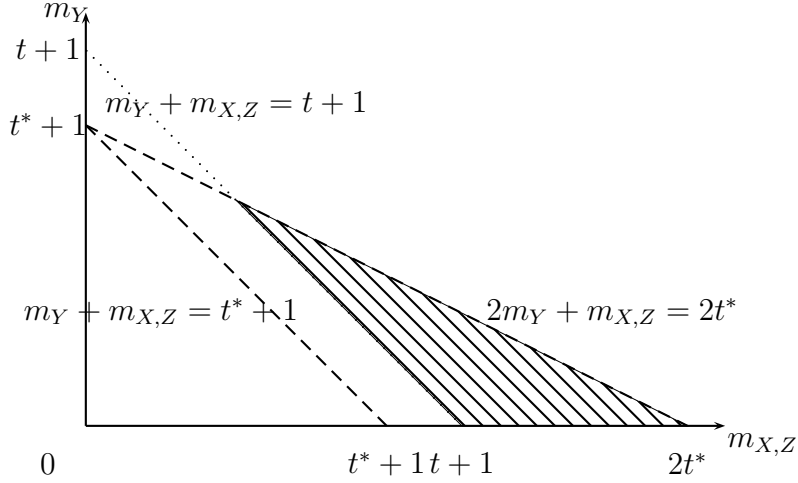

Let $E=i^{c}M_1\otimes M_2\otimes \ldots \otimes M_n= X_{\alpha}Z_{\beta}$, $\alpha =( a_1, \ldots, a_n )$, $\beta =( b_1, \ldots, b_n )\in \mathbb{Z}_{2}^{n}$ ,  be a correctable error operator with $w(\alpha)+w(\beta)\leqslant 2t^*$.
Let $l=gw(\alpha, \beta)$, the weight of $E$, and then $0\leqslant l\leqslant 2t^*$.
Let $m_Y= w(\alpha \beta)$, the number of $M_j$'s equal to $Y$, and then $m_{X,Z}=l-m_Y$ represents the number of the $M_j$'s equal to $X$ or $Z$.
Since $w(\alpha)+w(\beta)=2m_Y+m_{X,Z}=m_Y+l$, we have
    $
         m_Y+l\leqslant 2t^*
    $
(note that $l=m_{X,Z}+m_Y$).
There are
    \[
        {n\choose l} {l\choose m_Y} 2^{l-m_Y}
    \]
correctable error operators for a certain $l$ and a certain $m_Y$ satisfying $0\leqslant m_Y \leqslant l$ and $0\leqslant m_Y+l\leqslant 2t^*$.
Summing $l$ from $t^*+1$ to $2t^*$ and summing $m_Y$ from $0$ to $2t^*-l$, we have additional
    \begin{align}
         \sum_{l=t^*+1}^{2t^*}\sum_{m_Y=0}^{ 2t^*-l} {n\choose l}{{l}\choose {m_Y}} 2^{l-m_Y} \label{eq:add}
    \end{align}
correctable error operators of weight $>t^*$, which can be a large amount!
Figure \ref{fig:add} illustrates the $m_{X,Z}-m_Y$ region of all additional correctable error operators of weight $l=m_{X,Z}+m_Y > t^*$ as the dashed triangle.

 On the other hand, for a given stabilizer group $\mathcal {S}$ with quantum error-correcting capability $t$, the classical minimum distance $d'$ of a check matrix of $\mathcal{S}$ can be used to determine the existence of additional correctable error operators of weight $>t$
and how many of them as stated in the following theorem and illustrated in the dashed triangle
 in Figure \ref{fig:add}.
\bt
    Let $t$ be the quantum error correcting capability of a stabilizer code $\mathcal{C}(\mathcal{S})$.
    Let $t^*$ be an estimate of $t$ by a check matrix of $\mathcal{S}$ as in Theorem \ref{thm:H}.
    If  $t<2t^*$ or $\lfloor \frac{d-1}{2}\rfloor < 2 \lfloor \frac{d'-1}{4}\rfloor$,
    then    we have additional
    \[
         \sum_{l=t+1}^{2t^*}\sum_{m_Y=0}^{ 2t^*-l} {n\choose l}{{l}\choose {m_Y}} 2^{l-m_Y}
    \]
correctable error operators of weight $>t$.
\label{thm:dc adds}
\et

Finally we give a slight improvement of Theorem \ref{thm:H}.
Define a matrix $H_Y\triangleq H_X + H_Z$ for a check matrix    $
H= \left(\begin{array}{c@{\vdashline}c} H_X  & H_Z
\end{array}\right)$ of $\mathcal{S}$. Let $\mathcal{C}_X,$ $
\mathcal{C}_Z,$ $ \mathcal{C}_Y$ be classical binary linear block
codes with  parity-check matrices $H_X,$ $ H_Z,$ $ H_Y$,
respectively. If error syndromes of all correctable error
operators of weight no more than $t^*$ are distinct, it is
necessary that  $\mathcal{C}_X, \mathcal{C}_Z, \mathcal{C}_Y$
have minimum distance $\geqslant 2t^*+1.$
\bc
    If $C_Y$ has minimum distance $\geqslant 2t^*+1$, the condition in {Theorem \ref{thm:H}}  can be reduced to $d'\geq 4t^*$. \label{col:4t}
\ec
\begin{proof}
We only need to consider two distinct error operators $E_1=X_{\alpha_1 }Z_{\beta_1 }$ and $E_2=X_{\alpha_2 }Z_{\beta_2 }$ with
$\alpha_1 =( a_1, \ldots, a_n )$, $\beta_1=(b_1, \ldots, b_n)$, $gw(\alpha_1, \beta_1)= t^* $ and $\alpha_2 =( u_1, \ldots, u_n )$, $\beta_2 =(v_1, \ldots, v_n)$, $gw(\alpha_2, \beta_2)= t^* $
such that 
    \begin{align*}
        f_{\mathcal{S}}(E_1)+f_{\mathcal{S}}(E_2)=&  \sum_{i=1}^n (a_i+u_i)  f_{\mathcal{S}}(X_i) + \sum_{i=1}^n (b_i+v_i)  f_{\mathcal{S}}(Z_i)
    \end{align*}
 is a linear combination of $4t^*$ columns of $H$.
 Then we must have  $\alpha_1=\beta_1$, $\alpha_2=\beta_2$, $w(\alpha_1+\alpha_2)=2t^*$ and then
 \begin{align*}
        f_{\mathcal{S}}(E_1)+f_{\mathcal{S}}(E_2)=  \sum_{i=1}^n (a_i+u_i)  f_{\mathcal{S}}(Y_i),
    \end{align*}
which is a linear combination of $2t^*$ columns of $H_Y$.
Since $C_Y$ has minimum distance $\geqslant 2t^*+1$, we have
    \[
        f_{\mathcal{S}}(E_1)+f_{\mathcal{S}}(E_2)\neq 0.
    \]
Thus (\ref{4t}) holds for the extreme case---a linear combination of exact $4t^*$ terms.
\end{proof}


\subsection{Constructions of Check Matrices}
\label{sec:constructions-commutative-matrices}
Our construction of a check matrix of a stabilizer group needs a binary commutative parity-check matrix of even length.
We suggest three ways to establish commutative parity-check matrices by using classical constructions of new codes from old ones \cite{MS77}
such that the minimum distances of the resulted quantum codes can be determined from the corresponding classical binary linear block codes.

\noindent
{\bf Construction I:} Let $G_1, G_2$ be generator matrices of an $[n,k_1,d_1 ]$ and
an $[n,k_2,d_2 ]$ binary linear block codes $\mathcal{C}_1,
\mathcal{C}_2$, respectively, such that $k_1+k_2>n$. Let $H_1,H_2$
be parity-check matrices of $\mathcal{C}_1, \mathcal{C}_2$,
respectively. Then $ G'=\left[ \begin{array}{c|c}  G_1 & O \\O &
G_2  \end{array}\right]$ is a generator matrix of a
$[2n,k_1+k_2,d'=\min\{d_1,d_2\}]$ code with a parity-check matrix
$ H'=\left[ \begin{array}{c|c}  H_1 & O \\O &
H_2\end{array}\right]$. $H'$ is commutative if and only if $H_1
H_2^{T}= O$, i.e., the classical dual code of $\mathcal{C}_2$ is
a subcode of $\mathcal{C}_1$. In this way, by choosing $H=H'$ as
a check matrix of a stabilizer group $\mathcal{S}$ and
$G=G'\Lambda =\left[ \begin{array}{c|c}  O & G_1 \\G_2 & O
\end{array}\right]$ as a generator matrix of the symplectic dual
code $\bar{S}^{\bot *}$,
$\mathcal{C}(\mathcal{S})$ is an $[[n,k_1+k_2-n,d\ge d^*=\min\{d_1,d_2\}]]$ quantum code, 
where $d^*$ equals to the  minimum generalized weight of $\bar{S}^{\bot *}$, which corresponds to a vector of minimum weight in $C_1$ or in $C_2$.
Since $d^*=d'$, we have $\lfloor \frac{d^*-1}{2}\rfloor \geq 2\lfloor \frac{d'-1}{4}\rfloor$ and there is no additional correctable error operators of weight $>\lfloor \frac{d^*-1}{2}\rfloor$ guaranteed by Theorem \ref{thm:dc adds}.
Note that the construction of CSS codes is a special case of Construction I.

\noindent
{\bf Construction II:} Let $G_1, G_2$ be generator matrices of an  $[n,k,d_1 ]$  and an $[n,k,d_2 ]$ binary linear block codes  $\mathcal{C}_1, \mathcal{C}_2$, respectively.
Let $H_1,H_2$ be parity-check matrices of $\mathcal{C}_1, \mathcal{C}_2$, respectively.
Then $ G'=\left[ G_1 \right|\left. G_2\right]$ is a generator matrix of a $[2n,k,d'\geqslant d_1+d_2]$ binary linear block code $\mathcal{C}'$.
A parity-check matrix of $\mathcal{C}'$ is $H'=\left[  \begin{array}{cc} H_1 & O\\ O& H_2\\ A & B\end{array}   \right]$,
where $A,B$ are two matrices such that $G_1A^T+G_2B^T=O$ and the rank of $H'$ is $2n-k$, which is greater than $n$.
Since $H'$ has too many rows to be a check matrix,
we consider the code $\mathcal{C}'^{\bot}$, the classical dual code of $\mathcal{C}'$, instead.
If the matrix $G'$ is commutative, we choose $ H=G'=\left[ G_1 \right|\left. G_2 \right]$ as a check matrix  of a  stabilizer group $\mathcal{S}$. 
However, it remains to determine the classical minimum distance of  $\mathcal{C}'^{\bot}$.
%
%
 
\noindent
{\bf Construction III:} ($|u|u+v|$ construction) Let $G_1$ and $ G_2$ be generator matrices of an  $[n,k_1,d_1 ]$ and an $ [n,k_2,d_2 ]$ binary linear block codes $\mathcal{C}_1, \mathcal{C}_2$, respectively.
Then $ G'=\left[ \begin{array}{c|c}  G_1 & G_1 \\O & G_2 \end{array}\right]$ is a generator matrix of a $[2n,k_1+k_2,d'\geqslant \min\{d_1,d_2\}]$ code.
Let $H_1,H_2$ be parity-check matrices of $\mathcal{C}_1, \mathcal{C}_2$, respectively.
A parity-check matrix is $ H'=\left[ \begin{array}{c|c}  H_2 & H_2 \\H_1 & O\end{array}\right]$.
$H'$ is commutative if and only if $H_1 H_2^{T}= O$, i.e., the classical dual code of $\mathcal{C}_2$ is a subcode of $\mathcal{C}_1$.
In this case, we take $H'$ as a check matrix $H$ of a stabilizer group $\mathcal{S}$.
The minimum generalized weight of a generator matrix $G=G'\Lambda $ of $\bar{\mathcal S}^{\bot *}$ is $\min\{d_1,d_2\}$.
Hence the stabilizer code $\mathcal{C}(\mathcal{S})$ has parameters  $[[n,k_1+k_2-n,d\geqslant\min\{d_1,d_2\}]].$

When $\mathcal{C}_1$ is a subcode of $\mathcal{C}_2$, we consider the effect of a permutation matrix $P'= \begin{pmatrix} I & O\\ O & P\end{pmatrix}$ on the $|u|u+v|$ construction.
Let  $H''=H'P'= \begin{pmatrix} H_2 & H_2 P \\  H_1 & O \end{pmatrix}$.
$H''$ is commutative  if and only if $H_2PH_2^T=H_2P^TH_2^T$ and $H_2PH_1^T=O$.
In this case,  we take $H''$ as a check matrix $H$ of a  stabilizer group $\mathcal{S}$.
Then $G= G'P'\Lambda = \begin{pmatrix} G_1 P &G_1 \\ G_2P & O \end{pmatrix}$ is a generator matrix of $\bar{S}^{\bot *}$.
Consider a nonzero codeword
\[(c_1,c_2)= (u_1|u_2)\left(\begin{array}{c@{\vdashline}c} G_2P  & G_2  \\G_{1}P & O  \end{array}\right)=(u_1G_2P+u_2G_1P,u_1G_2)\] in $\bar{\mathcal S}^{\bot *},$
with a $k_1$-tuple $u_1$ and a $k_2$-tuple $u_2$, not both zero tuples.
It is clear that  $c_1\in \mathcal{C}_2P$, $c_2\in \mathcal{C}_2$ and $w(c_1)\geqslant d_2$, $w(c_2)\geqslant d_2$.
Then
\begin{align*}
    gw(c_1,c_2)&= w(c_1)+w(c_1)-w(c_1c_2)\\ 
                &= \frac{1}{2}\left( w(c_1)+w(c_2)+ (w(c_1)+w(c_2)-2w(c_1c_2))\right)\\
                &=\frac{1}{2}\left( w(c_1)+w(c_2)+ w(c_1+c_2)\right)\\
                &\geqslant d_2+\frac{w(c_1+c_2)}{2}.
\end{align*}
If \[\min_{c_1\neq \textbf{0}\ \mbox{\scriptsize or}\ c_2\neq \textbf{0}} \frac{1}{2}w(c_1+c_2)>0,\]
we can obtain a quantum code of a greater minimum distance.

\bt  {\label{thm:PaddD}}
    Let $\mathcal{C}_1$ be a subcode of $\mathcal{C}_2$ with $d_1>d_2$ and  let $H_1,H_2$ be parity-check matrices of $\mathcal{C}_1, \mathcal{C}_2$, respectively.
    Assume that $P$ be a  permutation matrix such that
    \begin{align}
    H_1PH_1^T=H_1P^TH_1^T, \ \ H_1PH_2^T=O  \label{eq:PCommu}
    \end{align}
    and
    \begin{align}
    \mathcal{C}_1P=\mathcal{C}_1, \ \ \mathcal{C}_2P=\mathcal{C}_2. \label{eq:CPfixed}
    \end{align}
    Let $G_1$ and $G_2=\left[ \begin{array}{c}  G_3 \\ G_1 \end{array}\right]$ be generator matrices of $\mathcal{C}_1$ and $\mathcal{C}_2$, respectively, such that
    \begin{align}
    u G_3P \neq u G_3  \mbox{ for any nonzero $(k_2-k_1)$-tuple $u$}.\label{eq:CPNotFiexd}
    \end{align}
    Then there is an $[[n,k_2+k_1-n, d\geqslant\min\{d_1,\lceil \frac{3d_2}{2}\rceil\}]]$ stabilizer code $\mathcal{C}(\mathcal{S})$ with a corresponding stabilizer group $\mathcal{S}$
    such that
    \[
    G= \left(\begin{array}{c@{\vdashline}c} G_2P  & G_2 \\  G_1P & O \end{array}\right)=\left(\begin{array}{c@{\vdashline}c} G_3P  & G_3 \\ G_1P & G_1 \\G_{1}P & O  \end{array}\right)
    \]
    is a generator matrix of $\bar{S}^{\bot *}$.
\et
\begin{proof}
A check matrix of $\mathcal{S}$ is
    \[
        H= \begin{pmatrix} H_1 & H_1 P \\  H_2 & O \end{pmatrix}
    \]
since Eq.~(\ref{eqn:commutative}) holds by (\ref{eq:PCommu}).
Consider a nonzero codeword
\[(c_1,c_2)= (u_1|u_2|u_3)\left(\begin{array}{c@{\vdashline}c} G_3P  & G_3 \\ G_1P & G_1 \\G_{1}P & O  \end{array}\right)=(u_1G_3P+ (u_2+u_3)G_1P,u_1G_3+u_2G_1)\] in $\bar{\mathcal S}^{\bot *},$
with two $k_1$-tuples $u_1,u_2$ and a $(k_2-k_1)$-tuple $u_3$, not all zero tuples.
It is clear that  $c_1\in \mathcal{C}_2P=\mathcal{C}_2, c_2\in \mathcal{C}_2$ and $w(c_1)\geqslant d_2, w(c_2)\geqslant d_2$.
If $u_1\neq 0$,
\begin{align*}
    c_1+c_2 &= u_1G_3P+ (u_2+u_3)G_1P + u_1G_3+u_2G_1\\
                &= (u_1G_3P+ u_1G_3)+ ((u_2+u_3)G_1P +u_2G_1).
\end{align*}
Since  $u_1G_3P+ u_1G_3  \in \mathcal{C}_2-\mathcal{C}_1$  by (\ref{eq:CPfixed}) and (\ref{eq:CPNotFiexd})
and $(u_2+u_3)G_1P +u_2G_1 \in \mathcal{C}_1$ by (\ref{eq:CPfixed}),
we have $c_1+c_2 \neq \textbf{0} $ in $ \mathcal{C}_2$ and $w(c_1+c_2)\geqslant d_2$.
Then
\[
    gw(c_1,c_2) 
    =\frac{1}{2}\left( w(c_1)+w(c_2)+ w(c_1+c_2)\right)  
    \geqslant \frac{3d_2}{2}.
\]
If $u_1= \textbf{0},$ we have $c_1 = (u_2+u_3)G_1P  \in\mathcal{C}_1$ and $c_2 = u_2G_1 \in\mathcal{C}_1$.
Hence
\begin{align*}
    gw(c_1,c_2)&= w(c_1)+w(c_2)-w(c_1c_2)    \geqslant d_1.
\end{align*}
Therefore, the minimum distance of the quantum code is \(d\geqslant \min\{d_1,\lceil \frac{3d_2}{2}\rceil\}.\)
\end{proof}

Further investigation on  the classical minimum distance may guarantee additional correctable error operators of weight greater than the quantum error-correcting capability by Theorem
\ref{thm:dc adds}.
A family of quantum Reed-Muller codes will be constructed by the $|u|u+v|$ construction
in Section \ref{sec:rm}.


\subsection{Existence of Commutative Parity-Check Matrices}
\label{sec:ExistenceOfCheckMatrices}

There is an important question: for a given $r\times (2n)$ parity-check matrix $H$, where $r<n$,
does there exist an effective permutation matrix $P$ such that $HP$ is commutative?

To answer this question, we run a simulation on a computer as follows.
Let $H= \left[ \begin{array}{c|c} I_{r\times r} &   B\end{array} \right]  $, where  $B$ is a randomly generated $r\times (2n-r)$ matrix.
Each element of $B$ is $1$ or $0$ with probability $p_1$ and $p_0=1-p_1$, respectively.
By exhaustive search with $(2n)!$ permutations for the case  $n=5$,
unfortunately, we found that there exists an $H$ which has no effective permutation matrix $P$ such that $HP$ is commutative.
However, there is a high probability that for a randomly generated matrix
$H= \left[ \begin{array}{c|c} I_{r\times r} &   B\end{array} \right]$,
there is an effective permutation matrix $P$ such that $HP$ is commutative.
Moreover, if $p_1<p_0$, the probability becomes higher.
This simulation suggests that parity-check matrices of classical LDPC codes may be transformed into legal check matrices by Theorem \ref{thm:H}.
However, it becomes extremely harder to verify this suggestion for $n\geq 8$ due to prohibitive
computing complexity.
The question of determining an effective permutation matrix for a parity-check matrix remains open.
The construction of quantum stabilizer codes can be converted to the construction of classical linear codes with commutative parity-check matrices.


\subsection{Asymptotic Coding Efficiency}
\label{sec:quantum-coding-bounds}

In this subsection, we will investigate the asymptotic coding efficiency of the construction of stabilizer codes as stated in Theorem \ref{thm:H} by assuming that among all $[n,k,d]$ binary linear block codes, there is at least one code with a parity-check matrix $H$ and an effective permutation matrix $P$ such that $HP$ is commutative.

\begin{figure}
\begin{center}
\includegraphics[height=8cm]{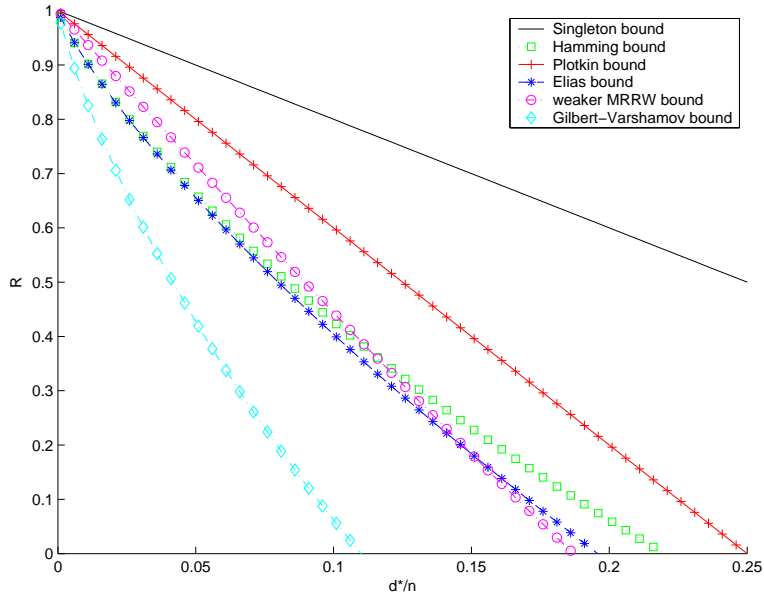}
\caption{Asymptotic coding bounds for the construction stated in Theorem \ref{thm:H}.} \label{fig:PQB}
\end{center}
\end{figure}

Suppose an $[[n,k, d \geq d^*=2 t^*+1 ]]$ stabilizer code with a check matrix is constructed by Theorem  \ref{thm:H} from a certain $[n'=2n,k'=k+n, d'\geq 4 t^*+1 ]$ classical linear block code, where $t^*=\lfloor \frac{d'-1}{4}\rfloor$.
Let $\alpha'=\limsup_{n'\rightarrow \infty } \frac{k'}{n'}$
 and $\alpha=\limsup_{n\rightarrow \infty } \frac{k}{n}$.
Since \[     \alpha'=\limsup_{n\rightarrow \infty } \frac{k+n}{2n}=\frac{1}{2}+\frac{1}{2}\alpha,\]
we have
\begin{align}
      \alpha =2\alpha'-1. \label{eq:alpha}
\end{align}
Let $\delta'=\frac{d'}{n'}$, $\delta^*=\frac{d^*}{n}$ and  $\delta=\frac{d}{n}$.
Since $\lfloor\frac{d'-1}{4}\rfloor \leq \frac{d'-1}{4}\leq \lfloor\frac{d'-1}{4}\rfloor+1$,
we have $t^* \leq \frac{d'-1}{4}\leq t^*+1$ and then
$2d^*-1 \leq d'\leq 2d^*+3$.
Thus we have $\frac{2d^*-1}{2n} \leq \delta' \leq \frac{2d^*+3}{2n}$ and then
$\delta^*-\frac{1}{2n} \leq \delta' \leq \delta^*+\frac{3}{2n}$.
Thus for sufficiently large $n$, we have
\begin{align}
     \delta^* \simeq \delta'. \label{eq:delta}
\end{align}
It is obvious that
\begin{align}
    \delta \geq \delta^*=\delta '. \label{eq:delta2}
\end{align}
From \cite{Lint82}, the classical Hamming bound says that 
        \[\alpha'(\delta') \leq 1-H_2(\frac{1}{2}\delta'),\]
        where $H_2(x)=-x\log_2(x)-(1-x)\log_2(1-x)$.
By (\ref{eq:alpha}) and (\ref{eq:delta}), we have a corresponding quantum Hamming bound of
the code construction in Theorem \ref{thm:H}, which is
\begin{align}
    \alpha(\delta^*) \leq 1- 2 H_2(\frac{1}{2}\delta^*).
\end{align}
The classical Plotkin Bound says that 
\begin{align*}
    \left.%
    \begin{array}{ll}
        \alpha'(\delta')\leq 1-2\delta', & \hbox{if $ 0\leq \delta' \leq \frac{1}{2}$,} \\
        \alpha'(\delta')=0, & \hbox{if $\frac{1}{2}< \delta' \leq 1$,}
    \end{array}%
    \right.
\end{align*}
and by (\ref{eq:alpha}) and (\ref{eq:delta}), the corresponding quantum Plotkin bound
of the code construction in Theorem \ref{thm:H} is
\begin{align}
    \left.%
    \begin{array}{ll}
        \alpha(\delta^*)\leq 1-4\delta^*, & \hbox{if $ 0\leq \delta^* \leq \frac{1}{4}$;} \\ \label{eq:plotkin}
        \alpha(\delta^*)=0, & \hbox{if $\frac{1}{4}< \delta^* \leq 1$.}
    \end{array}%
    \right.
\end{align}
The classical Elias Bound says that
\begin{align*}
    \left.%
    \begin{array}{ll}
        \alpha'(\delta')\leq 1-H_2(\frac{1}{2}-\sqrt{\frac{1}{2}(\frac{1}{2}-\delta')}), & \hbox{if $ 0\leq \delta' \leq \frac{1}{2}$,} \\
        \alpha'(\delta')=0, & \hbox{if $\frac{1}{2}< \delta' \leq 1$,}
    \end{array}%
    \right.
\end{align*}
and by (\ref{eq:alpha}) and (\ref{eq:delta}), the corresponding quantum Elias bound is
\begin{align}
    \left.%
    \begin{array}{ll}
    \alpha(\delta^*)\leq 1-2H_2(\frac{1}{2}-\sqrt{\frac{1}{2}(\frac{1}{2}-\delta^*)}), & \hbox{if $ 0\leq \delta^* \leq \frac{1}{2}$;} \\
    \alpha(\delta^*)=0, & \hbox{if $\frac{1}{2}< \delta^* \leq 1$.}
    \end{array}%
    \right.
\end{align}
The classical weaker McEliece-Rodemich-Rumsey-Welch (MRRW) bound says that
\begin{align*}
    \alpha'(\delta') \leq H_2(\frac{1}{2}-\sqrt{\delta'(1-\delta')})
\end{align*}
and by (\ref{eq:alpha}) and (\ref{eq:delta}), the corresponding weaker quantum MRRW bound is
\begin{align}
    \left.%
    \begin{array}{ll}
    \alpha(\delta^*) \leq 2 H_2(\frac{1}{2}-\sqrt{\delta^*(1-\delta^*)})-1, & \hbox{if $ \frac{1}{2}\leq H_2(\frac{1}{2}-\sqrt{\delta^*(1-\delta^*)}) \leq 1$;} \\
    \alpha(\delta^*)=0, & \hbox{if $0 \leq H_2(\frac{1}{2}-\sqrt{\delta^*(1-\delta^*)}) \leq \frac{1}{2}$.}
    \end{array}%
    \right.
\end{align}
%
%
The classical singleton bound says that
\begin{align*}
    \alpha'(\delta')\leq 1- \delta',
\end{align*}
and by (\ref{eq:alpha}) and (\ref{eq:delta}), the corresponding quantum singleton bound is
\begin{align}
    \alpha(\delta^*)\leq 1- 2\delta^*. \label{eq:singleton}
\end{align}
The classical Gilbert-Varshamov bound says that 
\begin{align*}
    \alpha'(\delta') \geq 1-H_2(\delta'),
\end{align*}
and by (\ref{eq:alpha}) and (\ref{eq:delta}), the corresponding quantum Gilbert-Varshamov bound is
\begin{align}
    \alpha(\delta^*) \geq 1-2H_2(\delta^*).
    \label{eq:Thm2-GVbound}
\end{align}
The above asymptotic bounds for the stabilizer code construction in Theorem \ref{thm:H}
are depicted in Fig. \ref{fig:PQB}.

\begin{figure}
\begin{center}
\includegraphics[height=8cm]{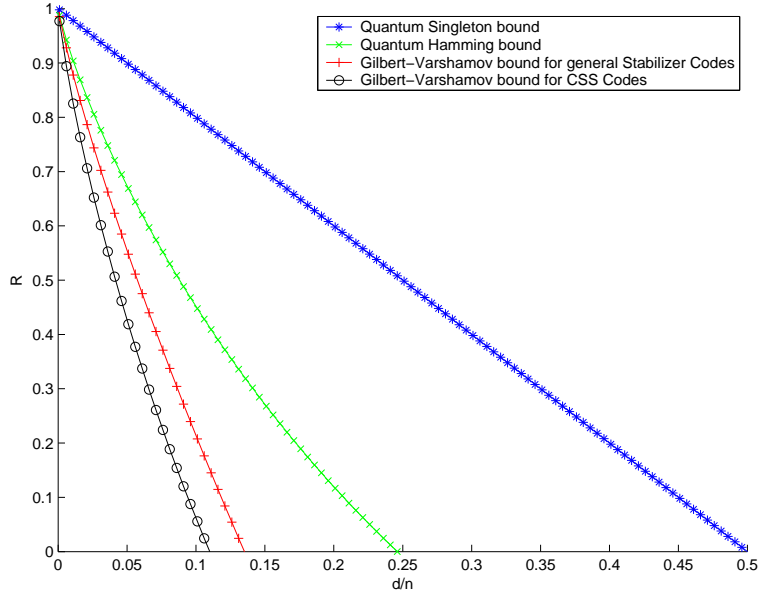}
\caption{Known quantum bounds in the literature.} \label{fig:KQB}.
\end{center}
\end{figure}

We next compare these asymptotic bounds with known bounds of quantum codes in the literature.
In \cite{EM96,Got96}, the quantum Hamming bound says that for an $[[n,k,d\ge d=2t+1]]$ quantum code,
    \begin{align*}
        2^k \sum_{i=0}^{t} 3^i {n \choose i} \leqslant 2^n,
    \end{align*}
and the asymptotic form is
        \[\frac{k}{n} \leq 1-\frac{t}{n}\log_{2}3-H_2(\frac{t}{n}),\]
or
\begin{align}
    \alpha(\delta) \leq 1-\frac{1}{2}\delta \log_{2}3-H_2(\frac{1}{2}\delta).
\end{align}
The quantum singleton bound \cite{KL97,Pre98} says that
for an $[[n,k,d]]$ quantum code,
    \[
        n-k\geq 2d-2,
    \]
or
\begin{align}
    \alpha(\delta^*)\leq 1- 2\delta. \label{eq:Qsingleton}
\end{align}
The Gilbert-Varshamov bound for a general quantum stabilizer codes, proved in Theorem 2 in \cite{CRSS97}, says that an $[[n,k,d=2t+1]]$ stabilizer code exists if
        \[
             \frac{k}{n} \geq 1-\frac{2t}{n} \log_{2}3-H_2(\frac{2t}{n}),
        \]
or
\begin{align}
    \alpha(\delta)\geq 1- \delta \log_{2}3-H_2(\delta).
    \label{eq:GeneralStabilizer-GVbound}
\end{align}
The Gilbert-Varshamov bound for CSS codes, proved in Section V in \cite{CS96}, says that an $[[n,k,d=2t+1]]$ CSS code exists if
        \[
            \frac{k}{n} \geq 1-2H_2(\frac{2t}{n}),
        \]
or
\begin{align}
    \alpha(\delta)\geq 1- 2H_2(\delta).
    \label{eq:CSS-GVbound}
\end{align}
The above known quantum bounds in the literature are depicted in Fig. \ref{fig:KQB}.
It can be seen that the two singleton bounds (\ref{eq:singleton}) and (\ref{eq:Qsingleton})
for the stabilizer code construction in Theorem \ref{thm:H} and for the general quantum codes,
respectively, are exactly the same.
And the two Gilbert-Varshamov bounds (\ref{eq:Thm2-GVbound}) and (\ref{eq:CSS-GVbound})
for the stabilizer code construction in Theorem \ref{thm:H} and for CSS codes are also exactly the same.
The Gilbert-Varshamov bounds (\ref{eq:GeneralStabilizer-GVbound}) for general stabilizer codes
is still better than the Gilbert-Varshamov bounds (\ref{eq:Thm2-GVbound})
for the stabilizer code construction in Theorem \ref{thm:H}.

%
%


\section{Quantum Reed-Muller Codes}
\label{sec:rm}
In this section, we will give a family of quantum
stabilizer codes from the parity-check matrices of Reed-Muller
codes by Theorem \ref{thm:H}.  
Parity-check matrices of Reed-Muller codes are commutative by the
$|u|u+v|$ construction. Permutation matrices that increase the
quantum minimum distance by Theorem \ref{thm:PaddD} are also
investigated.

\subsection{ Properties of Classical Reed-Muller Codes}
Reed-Muller codes are weakly self-dual codes and have simple but good structure properties \cite{MS77}.
A Reed-Muller code with two parameters $r,m$ is denoted by
$RM(r,m)$, $0\leq r\leq m$. This code is of length $2^m$ and $r$ is called its order.
Consider the following $(m+1)$ $2^m$-tuples
\[
\begin{array}{cccccccccccccccccccccc}
\mathbf{1} &=& (&1&1&1&1 & \cdots & 1&1&1&1 & 1&1&1&1 & \cdots & 1&1&1&1&),\\
       v_1 &=& (&0&1&0&1 & \cdots & 0&1&0&1 & 0&1&0&1 & \cdots & 0&1&0&1&),\\
       v_2 &=& (&0&0&1&1 & \cdots & 0&0&1&1 & 0&0&1&1 & \cdots & 0&0&1&1&),\\
      &\vdots&  & & & &\vdots  & \ddots & \vdots & & & & & & & \vdots & \ddots & \vdots & & & &\\
       v_m &=& (&0&0&0&0 & \cdots & 0&0&0&0 & 1&1&1&1 & \cdots & 1&1&1&1&).
\end{array}
\]
Then $RM{(r,m)}$ is generated by
    \begin{align*} &\mbox{(degree 0) }\mathbf{1},\\
                    &\mbox{(degree 1) } v_1, \ldots, v_m, \\
                    &\mbox{(degree 2) } v_1 v_2, \ldots, v_{m-1}v_m, \\
                    & \quad \quad \quad \vdots \\
                    &\mbox{(degree $r$) } v_1v_2\cdots v_{r}, \ldots  ,v_{m-r+1}v_{m-r+2}\cdots v_{m},
    \end{align*}
where the product of the $v_i$'s means the bitwise AND of the $v_i$'s and the degree means the number of $v_i$'s appearing in the product.
There are several properties of $RM(r,m)$ which can be derived directly from its construction \cite{MS77}.
The dimension of $RM(r,m)$ is
    \[
        k= \sum_{i=0}^{r}{m\choose i}
    \]
and the minimum distance of $RM(r,m)$ is
    \[
        d=2^{m-r}.
    \]
The dual code of $RM(r,m)$ is
    \[
        RM(m-r-1,m)
    \]
for $0\leq r \leq m$, where $RM(-1,m)\triangleq \{\textbf{0}\}$.
Let $G_{(r,m)}$ denote a generator matrix of $RM(r,m)$.
Reed-Muller code $RM(r,m+1)$ can be obtained from $RM(r,m)$ and $RM(r-1,m)$ by using the $|u|u+v|$ construction.
A generator matrix of $RM(r,m+1)$ is
\begin{align}
    G_{(r,m+1)} =\begin{pmatrix} G_{(r,m)} &G_{(r,m)} \\ O & G_{(r-1,m)}\end{pmatrix}.\label{eq:rm}
\end{align}
Since $RM(m-r-1,m)$ is the dual code of $RM(r,m)$, a parity-check matrix of $RM(r,m)$ is $G_{(m-r-1,m)}$ and
    \[
        G_{(r,m)}G_{(m-r-1,m)}^T =O.
    \]
For convenience, the orthogonality of Reed-Muller codes can be remarked in the following lemma.
\bl
\label{le:org rm}
    For $r+s\leq m-1$ and $m\geq 1$, $G_{(r,m)}$  and $G_{(s,m)}$ are orthogonal.

\el

Next we will consider the relation between the commutativity of $G_{(r,m)}$ and the parameters $r,m$.
We first consider the case when the permutation matrix $P$ is an identity matrix $I$ in (\ref{eq:HP}).
It is trivial that $G_{(0,m)}$ is commutative for all $m\geq 1$ since it is an all 1 vector.
In general, we have the following lemma.
\bl
    For $r\leq \lfloor \frac{m}{2}\rfloor$ and $m\geq 1$, $G_{(r,m+1)}$ is commutative.
\el
\begin{proof}
    By (\ref{eq:rm}), we have
    \begin{align*}
        G_{(r,m+1)}= \begin{pmatrix} G_{(r,m)} &G_{(r,m)} \\ O & G_{(r-1,m)} \end{pmatrix}.
    \end{align*}
    Since $r+(r-1)\leqslant m-1$, by Lemma \ref{le:org rm}, we have
    \[
        G_{(r,m)}G_{(r-1,m)}^T =O.
    \]
Then the commutativity of $G(r,m+1)$ can be verified as
    \begin{align*}
        G_{(r,m+1)}\Lambda G_{(r,m+1)}^T
        =& \begin{pmatrix} G_{(r,m)} &G_{(r,m)} \\
                            O & G_{(r-1,m)}\end{pmatrix}
            \begin{pmatrix} O & I \\
                            I & O \end{pmatrix}
            \begin{pmatrix} G_{(r,m)}^T& O \\
                            G_{(r,m)}^T  & G_{(r-1,m)}^T\end{pmatrix}\\
        =& \begin{pmatrix} G_{(r,m)} &G_{(r,m)} \\
                            O & G_{(r-1,m)}\end{pmatrix}
            \begin{pmatrix} G_{(r,m)}^T& G_{(r-1,m-1)}^T \\
                            G_{(r,m)}^T  & O\end{pmatrix}\\
        =& \begin{pmatrix} G_{(r,m)}G_{(r,m)}^T+ G_{(r,m)}G_{(r,m)}^T& G_{(r,m)}G_{(r-1,m)}^T \\
                            G_{(r-1,m)}G_{(r,m)}^T & O\end{pmatrix}\\
        =& \quad O.
    \end{align*}
\end{proof}

\subsection{Quantum Reed-Muller Codes from Parity-Check Matrices}
Now we will present a class of stabilizer codes derived from Reed-Muller codes
by Theorem \ref{thm:H}.

Let $m\geq 2r$ and let $H=G_{(r,m+1)}$ be a  check matrix of a
stabilizer group $S$. Then the stabilizer code
$\mathcal{C}(\mathcal{S})$ is a quantum code of {length $n=2^m$}
and dimension
    \[
        k= 2^m-\sum_{i=0}^r {{m+1}\choose i}.
    \]
The classical minimum distance of the parity-check matrix $G_{(r,m+1)}$ is
    \[
        d'=2^{(m+1)-((m+1)-r-1)}=2^{r+1}.
    \]
Thus the quantum error-correcting capability $t$ of $\mathcal{C}(\mathcal{S})$  is lower bound by
    \begin{align}
        t'=\lfloor \frac{d'-1}{4} \rfloor = 2^{r-1}-1, \label{eq:d c rm}
    \end{align}
by Theorem \ref{thm:H}.
Therefore, this quantum code $\mathcal{C}(\mathcal{S})$ is able to correct at least ($2^{r-1}-1$) qubit errors provided that $r\geq 1.$
The quantum minimum distance $d$ of  $\mathcal{C}(\mathcal{S})$ is lower-bounded by $2t'+1=2^r-1$.
On the other hand,
the symplectic dual  $\bar{\mathcal{S}}^{\bot}$ has { a generator matrix $G=G_{(m-r,m+1)}\Lambda $ which generates the same code as the generator matrix $G_{(m-r,m+1)}$}
and its generalized Hamming weight is $2^r.$

We have the following theorem.
\bt
    The parity-check matrix of a classical Reed-Muller code $RM(r,m+1)$ in (\ref{eq:rm}) with $m\geq 2r$ and $r\geq 1$ is a check matrix of a  $[[2^m,2^m-\sum_{i=0}^r  {{m+1}\choose i},  2^r]]$ quantum stabilizer code.
    \label{thm:QRM}
\et
Since the quantum error correcting capability $t=2^{r-1}-1$ equals to the lower bound $t'=2^{r-1}-1$ by Theorem \ref{thm:H},
this quantum code will have additional correctable error operators of weight $>t$.

Since Reed-Muller codes $RM(r,m)$ with $2r+1\leqslant m$  are  weakly self-dual codes,
by Lemma \ref{le:org rm}, we can use them to construct CSS codes.
Take $C_1=RM(r_1,m)$ with minimum distance $2^{m-r_1}.$
Then choose $C_2=RM(r_2,m)$, a subcode of $C_1$, with $ r_2 < r_1.$
The dual code of $C_2$ is $C_2^{\bot } =RM(m-r_2-1,m)$ with minimum distance $2^{m-{m-r_2-1}}=2^{r_2+1}.$
By CSS construction, we obtain a quantum code with parameters $[[2^{m},\sum_{i=r_2+1}^{r_1} {m\choose i}, \geq \min \{ 2^{m-r_1},2^{r_2+1}\}]]$.
For the best efficiency, we take $ r_2+1=m-r_1$.
Let $r=r_2+1$. Then $2r\le r_2+r_1+1=m.$
We now construct a  CSS code with parameters
    \begin{align}
        [[2^m,2^m-2\sum_{i=0}^{r-1}{m\choose i},2^{r}]]. \label{eq:RM-CSS}
    \end{align}
Comparing the dimension of a CSS code in (\ref{eq:RM-CSS}) with that of a quantum code in Theorem \ref{thm:QRM}, both having the same length $2^m$ and the same minimum distance $2^r$,
    \[
        (2^m-2\sum_{i=0}^{r-1}{m\choose i})-(2^m-\sum_{i=0}^{r}  {{m+1}\choose i})
        = {{m}\choose {r}},
    \]
we find that the CSS construction has a higher efficiency.
However, the construction in Theorem \ref{thm:QRM} gives us additional correctable error operators of weight $>t=2^{r-1}-1$.
In Table \ref{tb:comparision with CSS}, we list the number of additional correctable error operators for the stabilizer codes constructed by Theorem \ref{thm:H} from Reed-Muller codes with parameters
$(m,r) = (5,2), (6,3), (7,3)$.

\begin{table}[h]
\caption{A list of numbers of additional correctable error operators for the stabilizer codes
constructed by Theorem \ref{thm:H} from Reed-Muller codes with parameters
$(m,r) = (5,2), (6,3), (7.3)$.}
\[
         \begin{tabular}{|c|c|c|c|}
        \hline
        $n=2^m$ &   32  &   64   &  128    \\
        $t$    &    1 &   3 &   3\\
        \hline
        \# of additional correctable error operators    &    1984  &   5.99E+09 &  3.87E+11\\
        \# of original correctable error operators    &    97 &   1.14E+06 &   9.29E+06\\
        \# additional/ \# original    &    20.45 &   5.24E+03 &   4.16E+04\\
        \hline
        $m$    &    5 &   6 &  7\\
        $r$    &    2 &   3 &   3\\
        \hline
        dimension deficit ${m \choose {r}}$  relative to the CSS code  &    10 &   20 &   35\\
         &&&\\
        \hline
        \end{tabular}
\]
\label{tb:comparision with CSS}
\end{table}

On the other hand, when comparing the efficiency of
the $[[2^m,2^m-\sum_{i=0}^{r-1}  {{m+1}\choose i},  2^{r-1}]]$ stabilizer code constructed by Theorem \ref{thm:H} from a Reed-Muller code with that of the
$[[2^m,2^m-2\sum_{i=0}^{r-1}{m\choose i},2^{r}]]$ CSS code,
the former code has a surplus ${{m}\choose {i-1}}$ in dimension,
while the minimum distance of the former quantum code is only half of that of the latter CSS code.
However, the former code has a lot of additional correctable error operators
which will strengthen the error performance of the former code.

%
%

%
%

\subsection{Permutations Which Increase the Minimum Distance }
We find that
%
%
if we multiply $G_{(1,m+1)}$ by the permutation matrix $P'= \begin{pmatrix} I & O\\ O & P\end{pmatrix}$ with $P$ being the permutation matrix used in \cite{Got96},
a stabilizer group ${\mathcal{S}}$ with a check matrix $H=G_{(1,m+1)}P'$ will give a quantum stabilizer code  $\mathcal{C}(\mathcal{S})$ with parameters $[[2^m, 2^m-m-2,  3]]$,
which are the same as those constructed in \cite{Got96}.
For example, when $m=3$,
    \[
        P=\begin{pmatrix}
        0\quad 1\quad 0\quad 0\quad 0\quad 0 \quad 0\quad 0\\
        0\quad 0\quad 0\quad 1\quad 0\quad 0 \quad 0\quad 0\\
        0\quad 0\quad 0\quad 0\quad 1\quad 0 \quad 0\quad 0\\
        0\quad 0\quad 0\quad 0\quad 0\quad 0 \quad 1\quad 0\\
        0\quad 0\quad 0\quad 0\quad 0\quad 0 \quad 0\quad 1\\
        0\quad 0\quad 0\quad 0\quad 0\quad 1 \quad 0\quad 0\\
        0\quad 0\quad 1\quad 0\quad 0\quad 0 \quad 0\quad 0\\
        1\quad 0\quad 0\quad 0\quad 0\quad 0 \quad 0\quad 0
        \end{pmatrix} .
    \]
%
%
There are many other permutation matrices that will work by Theorem 10 in \cite{CRSS98},
which means that a column permutation on a parity-check matrix may give a stabilizer code $\mathcal{C}(\mathcal{S})$ with higher quantum error-correcting capability.
However, it is in general hard to find such a permutation matrix that the commutative condition still holds after the column permutation of the parity-check matrix.

We now investigate the effect of permutation matrices on generator matrices of Reed-Muller codes.
\bt
    Let $P'= \begin{pmatrix} I & O\\ O & P\end{pmatrix}$,
     where $P$ is a permutation matrix such  that
     the assumptions (\ref{eq:PCommu}), (\ref{eq:CPfixed}), and (\ref{eq:CPNotFiexd}) in Theorem \ref{thm:PaddD} hold with $\mathcal{C}_1=RM(m-r-1,m)$, $\mathcal{C}_2=RM(m-r,m)$ and $d_1=2^{r+1}$, $d_2=2^{r}$.
    Then for $m\geq 2r$ and $r\geq 1$, the quantum stabilizer code $\mathcal{C}(\mathcal{S})$ with a check matrix $H=G_{(r,m+1)}P'$ will have parameters
 $[[2^m,2^m-\sum_{i=0}^r  {{m+1}\choose i},  \geq 2^r+ 2^{r-1}]]$.
 In addition, $\mathcal{C}(\mathcal{S})$ will have additional correctable error operators if $r\geq 3$.   \label{thm:PQRM}
\et
\begin{proof}
By Theorem \ref{thm:PaddD}, the minimum distance of $\mathcal{C}(\mathcal{S})$  is at least
 \[
    \min\{ d_1, \lceil \frac{3d_2}{2} \rceil\}=\min\{ 2^{r+1}, \lceil \frac{3\times 2^{r}}{2} \rceil\}= 2^r+ 2^{r-1}.
 \]
Note that the classical minimum distance of the parity-check matrix $H=G_{(r,m+1)}P'$ remains unchanged after a column permutation.
By (\ref{eq:d c rm}), $t'=2^{r-1}-1$.
For $d=2^r+2^{r-1}$, the error-correcting capability of the quantum code  $\mathcal{C}(\mathcal{S})$ is $t=\lfloor \frac{d-1}{2} \rfloor = \frac{3}{2}2^{r-1}-1$.
Then we have $2t'-t=2^{r-2}-1 >0$
if $r\geq3.$
Thus by Theorem \ref{thm:dc adds}, the quantum code  $\mathcal{C}(\mathcal{S})$ will have additional correctable error operators of weight $>t= \frac{3}{2}2^{r-1}-1$  if $r\geq 3$.
\end{proof}

In \cite{Ste98}, Steane gave a class of quantum Reed-Muller codes with parameters $[[2^m,2^m-\sum_{i=0}^r  {{m+1}\choose i},  \geq 2^r+ 2^{r-1}]]$ as given by Theorem \ref{thm:uuv}.
If there exists a permutation matrix $P'$ satisfying the assumptions in Theorem \ref{thm:PQRM},
a stabilizer group ${\mathcal{S}}$ with a check matrix $H=G_{(r,m+1)}P'$, $r\geq 1$,
will give a quantum stabilizer code  $\mathcal{C}(\mathcal{S})$ having the same parameters  as those in \cite{Ste98} but having additional correctable error operators.

We now give an effective permutation matrix for $G_{(1,m)}$.
Let
\[
    T=\left[
            \begin{array}{c}
                I_{\frac{n}{2}} \otimes \begin{bmatrix}1 & 0 \end{bmatrix} \\
                I_{\frac{n}{2}} \otimes \begin{bmatrix}0 & 1 \end{bmatrix} \\
            \end{array}
        \right],
\]
and
\[
    Q=\left[
            \begin{array}{cc}
                I_{\frac{n}{2}} & O \\ O & I_{\frac{n}{4}} \otimes \begin{bmatrix}0 & 1\\1  & 0\end{bmatrix} \\
            \end{array}
        \right],
\]
where $n=2^m$.
For example, when $n=2^3=8$,
    \[
        T=\begin{pmatrix}
        1\quad 0\quad 0\quad 0\quad \quad 0\quad 0 \quad 0\quad 0\\
        0\quad 0\quad 1\quad 0\quad \quad 0\quad 0 \quad 0\quad 0\\
        0\quad 0\quad 0\quad 0\quad \quad 1\quad 0 \quad 0\quad 0\\
        0\quad 0\quad 0\quad 0\quad \quad 0\quad 0 \quad 1\quad 0\\
        0\quad 1\quad 0\quad 0\quad \quad 0\quad 0 \quad 0\quad 0\\
        0\quad 0\quad 0\quad 1\quad \quad 0\quad 0 \quad 0\quad 0\\
        0\quad 0\quad 0\quad 0\quad \quad 0\quad 1 \quad 0\quad 0\\
        0\quad 0\quad 0\quad 0\quad \quad 0\quad 0 \quad 0\quad 1
        \end{pmatrix} ,
    \]
    \[
        Q=\begin{pmatrix}
        1\quad 0\quad 0\quad 0\quad \quad 0\quad 0 \quad 0\quad 0\\
        0\quad 1\quad 0\quad 0\quad \quad 0\quad 0 \quad 0\quad 0\\
        0\quad 0\quad 1\quad 0\quad \quad 0\quad 0 \quad 0\quad 0\\
        0\quad 0\quad 0\quad 1\quad \quad 0\quad 0 \quad 0\quad 0\\
        0\quad 0\quad 0\quad 0\quad \quad 0\quad 1 \quad 0\quad 0\\
        0\quad 0\quad 0\quad 0\quad \quad 1\quad 0 \quad 0\quad 0\\
        0\quad 0\quad 0\quad 0\quad \quad 0\quad 0 \quad 0\quad 1\\
        0\quad 0\quad 0\quad 0\quad \quad 0\quad 0 \quad 1\quad 0
        \end{pmatrix} .
    \]
It can be verified that $v_i T = v_{i+1}$ for $1\leq i \leq (m-1)$ and $v_m T= v_1$.
Thus
    \[
        P=T Q
    \]
is a permutation matrix such that $v_i P= v_{i+1}$ for $1\leq i \leq (m-1)$ and $v_m P= v_1+v_m $.
\bt
      The permutation matrix $P=T Q$ is a permutation matrix
     such  that the assumptions (\ref{eq:PCommu}), (\ref{eq:CPfixed}), and (\ref{eq:CPNotFiexd}) in Theorem \ref{thm:PaddD} hold with $\mathcal{C}_1=RM(m-2,m)$, $\mathcal{C}_2=RM(m-1,m)$ and $d_1=4, d_2=2$ for $m\geq 2$.
    Then a quantum stabilizer code $\mathcal{C}(\mathcal{S})$ with a check matrix $H=G_{(1,m+1)}P'$ will have parameters
 $[[2^m,2^m-m-2,  \geq 3]]$.
\et
\begin{proof}
$H_1=G_{(1,m)}$ and  $H_2=G_{(0,m)}$ are parity-check matrices of $\mathcal{C}_1=RM(m-2,m)$ and $\mathcal{C}_2=RM(m-1,m)$, respectively.
It can be verified that $H_1PH_1^T=H_1P^TH_1^T$ and  $H_1PH_2^T=O$.
Denote $v_i P\triangleq v_i'$ for convenience.
It is obvious that
       $
            (v_{i_1}\cdots v_{i_l})P=v_{i_1}'\cdots v_{i_l}'
       $
    for any $l$.
%
%
Since $\mathcal{C}_2=RM(m-1,m)$ is generated by
        \[
            \mathbf{1}, v_1, \ldots, v_m,  v_1 v_2, \ldots, v_{m-1}v_m, \ldots,   v_1v_2\cdots v_{m-1}, \ldots, v_2v_3\cdots v_{m},
        \]
$\mathcal{C}_2P=RM(m-1,m)P$ is generated by
        \[
            \mathbf{1}P, v_1P, \ldots, v_mP,  (v_1 v_2)P, \ldots, (v_{m-1}v_m)P, \ldots,   (v_1v_2\cdots v_{m-1})P, \ldots, (v_2v_3\cdots v_{m})P
        \]
    or
            $\mathbf{1}, v_2, \ldots, v_m, v_1+v_m,$  $v_2 v_3, \ldots, v_mv_1+v_{m},$ $\ldots,$   $(v_2v_3\cdots v_{m})$, $\ldots$, $(v_3v_4\cdots v_{m}v_1)+(v_3v_4\cdots v_{m}).$
    It can be easily verified that $\mathcal{C}_2P= \mathcal{C}_2$.
    Similarly, we have $\mathcal{C}_1P= \mathcal{C}_1$.
    Two generator matrices of $\mathcal{C}_1=RM(m-2,m)$ and $\mathcal{C}_2=RM(m-1,m)$ are $G_1=G_{(m-2,m)}$ and  $G_2=\left[ \begin{array}{c}  G_3 \\ G_1 \end{array}\right]=G_{(m-1,m)}$, respectively.
    The row vectors of $G_3$ are $v_1v_2\cdots v_{m-1}$, $\ldots$, $v_2v_3\cdots v_{m}$, the $m$ generators of degree $m-1$.
    For convenience, we denote $ v_{i_1}\cdots v_{i_{m-1}}\triangleq w_j$ if $j$ is not equal to
    any $i_l$, $1\le l\le m-1$.
    Then the row vectors of  $G_3$ are $w_m, w_{m-1}, \ldots, w_2, w_{1}$
    and the row vectors of  $G_3P$ are $w_mP, w_{m-1}P, \ldots, w_2P, w_{1}P$ with
    \begin{align*}
        w_m P &= (v_1v_2\cdots v_{m-1})P=v_2v_3\cdots v_m=w_1, \\
        w_{m-1}P&= v_2v_3\cdots v_{m-1}v_m+v_2v_3\cdots v_{m-1}v_1=w_1+w_m \\
        w_{m-2}P&= v_2\cdots v_{m-2}v_{m}+v_2\cdots v_{m-2}v_{m}v_1=v_2\cdots v_{m-2}v_{m}+w_{m-1} \\
        & \vdots \\
        w_{2}P&= v_2v_4\cdots v_{m-1}v_mv_m+v_2v_4\cdots v_{m}v_1=v_2v_4\cdots v_m+w_3, \\
         w_1P&= v_3v_4\cdots v_{m-1}v_mv_m+v_3v_4\cdots v_{m}v_1=v_3v_4\cdots v_m+w_2.
    \end{align*}
    If $uG_3P=uG_3$ for some binary $m-$tuple $u=(a_1,\ldots, a_m)$,
    \begin{align*}
          \sum_{i=1}^m a_i w_i = \sum_{i=1}^m a_i w_iP.
    \end{align*}
    It can be verified that the above equation holds only if $a_1=a_2=\ldots =a_m=0$, that is, $u$ is a zero tuple.
    Therefore, $u G_3P \neq u G_3 \mbox{ for any nonzero $(k_2-k_1)-$tuple $u$}$.
    By Theorem \ref{thm:PQRM}, we have a $[[2^m,2^m-m-2,  \geq 3]]$ quantum stabilizer code $\mathcal{C}(\mathcal{S})$ with a check matrix $H=G_{(1,m+1)}P'$ for $m\geqslant 2$.
\end{proof}
There are other permutation matrices which will work by similar proofs.

For a general Reed-Muller code $RM(r,m)$ with minimum distance $d'=2^{m-r}$, $m\ge 2r$,
we have the following conjecture.
\bcj
    Either $P= T$ or $P=T Q$ is a  permutation matrix satisfying the assumptions in Theorem \ref{thm:PQRM} for all $r,m$ with $m\ge 2r$.
    \hfill $\Box$
\ecj

\section{Quantum Cyclic Codes}
\label{sec:cyclic}

\subsection{Quantum Circulant Codes}
%
%
Motivated by the five-qubit code in Eq.~(\ref{eqn:5qubitcode}), we use Construction II in Subsection \ref{sec:constructions-commutative-matrices} with a check matrix
$H=\left[H_X | H_Z\right]=[G_1| G_2]$,  where $G_1, G_2$ are generator matrices of two classical cyclic codes $\mathcal{C}_1,\mathcal{C}_2$, respectively.
This method is called the circulant construction and the generated quantum codes are called quantum circulant codes.

We arbitrarily choose two binary polynomials $g_1(x), g_2(x)$ as generator polynomials of  two classical cyclic codes $\mathcal{C}_1,\mathcal{C}_2$, respectively.
Then they cyclicly generate the matrices $H_X=G_1, H_Z=G_2$, respectively.
If $H$ is commutative, we then justify the rank of $H$ by transforming $H$ into the row-reduced-echelon form.
The minimum distances of these quantum codes are determined by computer search and
for small $n$, many good quantum circulant codes are found by this method and
the quantum circulant codes with parameters achieving the upper bound in Table III in \cite{CRSS98} are listed in Table \ref{tb:circulant}.
\begin{table}[bht]
\caption{ Some extremal quantum circulant codes.}
\[
         \begin{tabular}{|c|c|c|c|c|c|c|c|c|c|}
\hline
         $n\diagdown d\backslash k$ & 0 & 1& 2&3&4&5&6&7&8\\
\hline
         $5$ &3&3&&&&&&&\\
         $6$ &4&&2&2&2&&&&\\
         $7$ &3&3&&2&2&&&&\\
         $8$ &4&3&&&2&2&2&&\\
         $9$ &4&3&3&3&&&2&&\\
         $10$ &4&4&&&&2&2&&2\\
         $12$ &6&&4&4&&&&2&2\\
         $13$ &5&5&&&&&&&\\
         $14$ &6&5&5&4&4&&&3&3\\
         $15$ &6&5&5&5&4&4&4&3&3\\
         $16$ &6&6&&&&4&&&\\
         $17$ &7&7&&&&&&&4\\
         $18$ &&&6&&5&&&&\\
         $19$ &&7&&&&&&&\\
        \hline
        \end{tabular}
\]
 \label{tb:circulant}
\end{table}
It is conjectured that most of the best quantum codes can be obtained
by the circulant construction.
However, the minimum distance is hard to decide for large $n$ for the extremely high computing complexity.

\subsection{Quantum Quadrature-Residue Codes}
Another interesting fact of the five-qubit code is that the binary
bit patterns of the generator polynomials $g_1(x), g_2(x)$ are the
\emph{indicator vectors} of the quadratic-residues mod $5$ and the
non-residues, respectively.
Inspired by this fact, a suggestion for any prime $p$ of the form
$8m+5$ is given in
\cite{CRSS97} without a proof though there is no classical binary
quadratic-residue code for theses $p$'s.
In Subsubsection \ref{subsec:basicQR}, we first introduce the basics of quadratic
residues (please refer to \cite{MS77} for a detailed discussion) and then give several quantum quadratic-residue codes by the
CSS construction for prime numbers of the form $p= 8m\pm 1$.
We then use the circulant construction with the two indicator vectors
of the quadratic-residues and the
non-residues mod $p$ for prime numbers $p$ of the form $p= 4j+ 1$ in Subsubsection \ref{subsec:QQR}.
The minimum distance of the generated quantum codes will be calculated by computer search.

\subsubsection{Binary Quadrature-Residue Codes
and Their Applications}
\label{subsec:basicQR}
Let $p$ be an odd prime.
For a nonzero $j$, $1\le j\le p-1$, if $j=l^2 \mod p$
for some $l$, we say that $j$ is a quadratic-residue
$\mbox{mod } p$, otherwise $j$ is called a non-residue.
In particular, 0 is neither a residue nor a non-residue.
There are $\frac{1}{2}(p-1)$ quadratic-residues and  $\frac{1}{2}(p-1)$  non-residues mod $p$.
Let $Q$ denote the set of quadratic residues mod $p$ and $N$ denote the set of non-residues.
If $\rho$ is a primitive element of the field $GF(p)$, then $\rho^j \in Q $ if and only if $j$ is even.
Thus $Q$ is a cyclic group generated by $\rho ^2$.
It can be verified that $Q$ is a disjoint union of cyclotomic cosets mod $p$.
Let
\[q(x)= \prod_{r \in Q} (x-\alpha^r),\quad n(x)=\prod_{n\in N} (x-\alpha^n),\]
where $\alpha$ is a primitive $p$-th root of unity.
Then $x^p-1=(x-1)q(x)n(x).$
The quadratic-residue (QR) codes $\mathcal{Q,\bar{Q},N,\bar{N}}$ are
cyclic codes with generator polynomials $q(x),(x-1)q(x),n(x),(x-1)n(x)$, respectively.
The dimensions of $\mathcal{Q,N}$ are both $\frac{1}{2}(p+1)$ and the dimensions of $\mathcal{\bar{Q},\bar{N}}$ are both $\frac{1}{2}(p-1)$.
It is clear that $\mathcal{Q} \supseteq \mathcal{\bar{Q}} , \mathcal{N} \supseteq \mathcal{\bar{N}}$.

A binary polynomial $E(x)$ in the quotient ring $R_p=GF(2)[x]/(x^p-1)$
is an idempotent if $ E(x)= E(x)^2 =E(x^2)$ mod $x^p-1$. 
A binary cyclic code $\mathcal{C}=\langle g(x)\rangle $ of length $p$
contains a unique idempotent $E(x)$ such that $\mathcal{C}=\langle E(x)\rangle$.
Of course, $E(x)=p(x)g(x)$ for certain $p(x)$ .

Consider a prime $p$ such that $2\in Q.$
If $p=4j-1$, the $\alpha$ can be chosen such that the idempotents of
$\mathcal{Q,\bar{Q},N,\bar{N}}$ are
\[E_q(x)=\sum_{r\in Q} x^r, F_q(x)=1+\sum_{n\in N} x^n, E_n(x)=\sum_{n\in N} x^n, F_n(x)=1+\sum_{r\in Q} x^r.\]
In this case,
$\mathcal{Q}^{\bot} =\mathcal{\bar{Q}}$ and $\mathcal{N}^\bot =\mathcal{\bar{N}}$.
If $p=4j+1$, the $\alpha$ can be chosen such that the idempotents
of $\mathcal{Q,\bar{Q},N,\bar{N}}$ are
\[E_q(x)=1+\sum_{r\in Q} x^r, F_q(x)=\sum_{n\in N} x^n, E_n(x)=1+\sum_{n\in N} x^n, F_n(x)=\sum_{r\in Q} x^r.\]
In this case,
$\mathcal{Q}^{\bot} =\mathcal{\bar{N}}$ and $\mathcal{N}^\bot =\mathcal{\bar{Q}}$.

If $F_q(x)=\sum_{i=0}^{p-1} f_ix^i$, then a generator matrix of
$\mathcal{\bar{Q}}$ is
\[ \bar{G}= \begin{bmatrix} f_0& f_1 & \ldots & f_{p-1}\\f_{p-1}& f_0 & \ldots & f_{p-2}\\ \vdots & \vdots & \ddots & \vdots \\ f_1& f_2 & \ldots & f_{0}\\\end{bmatrix}\]
and a generator for $\mathcal{Q}$ is
\[ \left[  \begin{array}{c} {\bar{G}} \\ 11\ldots 1 \end{array}\right].\]
Similar results can be obtained for $\mathcal{N}$ and $\mathcal{\bar N}$.

It is obvious that the above dual pairs can be used in the CSS construction.
Taking $\mathcal{C}_1=\mathcal{Q},\mathcal{C}_2=\mathcal{\bar{Q}}$ or $\mathcal{C}_1=\mathcal{N},\mathcal{C}_2=\mathcal{\bar{N}}$ if $p=4j-1$, and
 $\mathcal{C}_1=\mathcal{Q},\mathcal{C}_2=\mathcal{\bar{N}}$ or $\mathcal{C}_1=\mathcal{N},\mathcal{C}_2=\mathcal{\bar{Q}}$ if $p=4j+1$,
 we can obtain a $[[p,1,d]]$ quantum code for certain $d$ in both cases.
By Theorem 1 and Theorem  23 in Ch.16 in \cite{MS77}, we have the following theorem similar to Theorem 40 and Theorem 41 in \cite{AKS06a} but of binary case.
\bt
If $p=8m\pm 1$, then there exists a $[[p,1,d]]$  CSS code with $d\geq \sqrt{p}$.
If $p=4j-1$, we can strengthen the bound to $d^2-d+1\geq p$.
\et
By collecting the minimum distance of various classical binary quadrature-residue codes in \cite{CS84,Boston99,TCL05}, we obtain Table \ref{tb:CCSQR}.

\begin{table}[bht]
\caption{Parameters of several $[[p,1,d]] $ CSS codes from classical binary quadrature-residue codes.}
\[
         \begin{tabular}{|c|c|c|c|c|c|c|c|}
        \hline
        $p$  &7&17&23&31 &41&47&71     \\
        \hline
        $d$  &3&5&7&7 &9&11&11     \\
        \hline
        \hline
        $p$  &73 &79&89&97&103&113 &137    \\
        \hline
        $d$  &13 &15&17&15&19&15 &21     \\
        \hline
        \end{tabular}
\]
 \label{tb:CCSQR}
\end{table}

Quadrature-residue codes can be extended by adding an overall parity-check bit so that the extended quadrature-residue codes
$\mathcal{\hat Q}$ and $\mathcal{\hat N}$ have the following relations:
\[ \mathcal{\hat Q}^{\bot} =\mathcal{\hat Q} \mbox{ and }  \mathcal{\hat N}^{\bot} =\mathcal{\hat N}, \mbox{ if $p=4j-1$}\]
 and
\[ \mathcal{\hat Q}^{\bot} =\mathcal{\hat N}, \mbox{ if $p=4j+1$. }\]
Similarly, the extended quadrature-residue codes can be used in the CSS construction and we have the following theorem by Theorem 8 in Ch.16 in \cite{MS77}.
\bt
If $p=8m\pm 1$, then there exists a $[[p+1,0,d]]$ CSS code for certain $d$.
If $p=4j-1$, $d\equiv 0 \mbox{ or } 3 \mbox{ mod }4$. If $p=4j+1$, $d$ is even.
\et

With Table 1(a) in \cite{Grassl00}, we have Table \ref{tb:CCSQR2}.

\begin{table}[bht]
\caption{Parameters of several $[[p+1,0,d]] $ CSS codes from classical binary extended quadrature-residue codes.}
\[
         \begin{tabular}{|c|c|c|c|c|c|c|c|}
        \hline
        $p$  &8&18&24&32 &42&48&72     \\
        \hline
        $d$  &4&6&8&8 &10&12&12     \\
        \hline
        \hline
        $p$  &74 &80&90&98&104&114 &128    \\
        \hline
        $d$  &14&16&18&16 &20&16&20     \\
        \hline
        \hline
        $p$  &138&152&168&192&194&200&    \\
        \hline
        $d$  &22&20&24&28&28 &32&     \\
        \hline

        \end{tabular}
\]
 \label{tb:CCSQR2}
\end{table}
The parameters in Table \ref{tb:CCSQR} are related to those in Table \ref{tb:CCSQR2} by Theorem 6 in \cite{CRSS98} in spite of the fact that
the entries are fewer in Table \ref{tb:CCSQR}.

\subsubsection{Quadrature Residues Related Quantum Circulant Codes}
\label{subsec:QQR}
Let $g_1(x)=\sum_{i=0}^{p-1}a_i x^i=\sum_{r\in Q} x^r$ and  $g_2(x)=\sum_{j=0}^{p-1}b_j x^j=\sum_{n\in N} x^n$ for certain prime number $p$.
These two binary polynomials are corresponding to the indicator vectors
of the quadratic-residues and the non-residues, respectively. For
example, when $p=13$,
    \[
        g_1(x)=x+x^3+x^4+x^9+x^{10}+x^{12}, \ g_2(x)=x^2+x^5+x^6+x^7+x^8+x^{11}
    \]
and the two corresponding indicator vectors are
    \[
        0101100001101 \ , \ 0010011110010 ,
    \]
respectively.
We begin with the following lemma to discuss our method for $p=4j+1$.
\bl
    For $p=4j+1$, the matrices \[ H_X= \begin{bmatrix} a_0& a_1 & \ldots & a_{p-1}\\a_{p-1}& a_0 & \ldots & a_{p-2}\\ \vdots & \vdots & \ddots & \vdots \\ a_1& a_2 & \ldots & a_{0}\\\end{bmatrix}\]
and
    \[ H_Z= \begin{bmatrix} b_0& b_1 & \ldots & b_{p-1}\\b_{p-1}& b_0 & \ldots & b_{p-2}\\ \vdots & \vdots & \ddots & \vdots \\ b_1& b_2 & \ldots & b_{0}\\\end{bmatrix}\]  are symmetric. \label{lemma:sym}
\el
\begin{proof}
We have $a_j=1$ if $j\in Q$ and  $a_j=0$, else, by the choice of $g_{1}(x)$.
For $p=4j+1$, $-1 \in Q$ and we have $a_j= a_{-j}=a_{p-j}$.
The element in the $i$-th row and the $j$-th column of $H_x$ is
$(H_X)_{ij}=a_{j-i\mod  p}$.
Since
$(H_X)_{ij}=a_{j-i\mod  p} =a_{-(j-i)\mod p}=a_{i-j\mod  p}=(H_X)_{ji}$,
the matrix $H_X$ is symmetric.
Similar reason holds for $H_Z$.
\end{proof}
\bt \label{thm:QR}
    For $p=4j+1$,  the matrix $H= \left[ H_X| H_Z\right]$  is commutative and has rank $p-1$.
    Then there is a $[[p,1,d]]$ quantum stabilizer code for a certain $d$.
\et
\begin{proof}
Let $ H_Y = H_X+H_Z$.
Equivalently, we consider a check matrix $\left[ H_Y| H_Z \right]$ instead of $\left[ H_X| H_Z \right]$.
Note that the rank of the check matrix remains to be determined.
Let $J_p$ denote the $p\times p$ all-1 matrix.
It is obvious that \[H_Y=\begin{bmatrix} 0& 1&1&\ldots &1\\ 1& 0&1&\ldots &1 \\1& 1&0&\ldots &1\\ \vdots& \vdots& \vdots &\ddots &\vdots \\1& 1&1&\ldots &0 \end{bmatrix}=J_p-I_p.\]
Denote the $i$-th row vectors of $H_Y$ and $H_Z$ by $\alpha_i$ and $\beta_i$, respectively.
Since $\alpha_i$ and $\beta_i$ are the indicator vectors both with $i$ right cyclic shifts,
we have $\alpha_i \cdot \beta_i=0$.
Note that $\alpha_i+\alpha_j=0\ldots 010\ldots 010\ldots 0$ with two 1's at the $i$-th and the $j$-th positions.
The commutative condition can be checked for any two rows as follows:
\begin{align*}
    \alpha_i\cdot \beta_j + \alpha_j\cdot \beta_i &=(\alpha_i+ \alpha_j)\cdot (\beta_i+\beta_j) \\
                                    &=(0\ldots 010\ldots 010\ldots 0 ) \cdot (\beta_i+\beta_j)\\
                                    &=(H_Z)_{i,i}+(H_Z)_{j,i}+(H_Z)_{i,j}+(H_Z)_{j,j} \ .
\end{align*}
Since $H_Z$ is symmetric by Lemma \ref{lemma:sym} and $(H_Z)_{i,i}=0$ for all $i$ by the construction,
the commutative condition holds.
%
%
For $H_Y =J_p-I_p$, the last row vector of $H_Y$ can be obtained from the summation of all the other row vectors.
Thus the rank of $H_Y $ is $p-1$.
Similarly, the ranks of $H_X $ and $H_Z $  are both $p-1$.
Hence the rank of the check matrix $H$ is $p-1$.
Thus the quantum stabilizer code with a check matrix $H$ has parameters $[[p,1,d]]$  for  a certain $d$.
\end{proof}
In \cite{CRSS97}, the quantum codes of $p=8m+5$ is a special case of above theorem.
It remained to determine the minimum distances of these codes.
Parameters of several quantum codes from Theorem \ref{thm:QR} are given in Table  \ref{tb:QR}.
The minimum distance of these codes is determined by computer search.
\begin{table}[bht]
\caption{Parameters of several $[[p,1,d]] $ quantum codes from Theorem \ref{thm:QR}.}
\[
         \begin{tabular}{|c|c|c|c|c|}
        \hline
        $p$  &5&13&17&29     \\
        \hline
        $d$  &3&5&5&11      \\
        \hline
        \end{tabular}
\]
 \label{tb:QR}
\end{table}
Note that the minimum distance of the quantum codes  in Table  \ref{tb:QR} with $p= 5, 13, 29$
achieves the upper bound in \cite{CRSS98}.
However, the minimum distance of a generic quantum code from Theorem \ref{thm:QR} is
not found for  the extremely high computing complexity.

\subsubsection{A Construction for Quantum Codes with $ k=1 $}
Inspired from the proof of the quantum quadratic residue codes, we give a construction of $[[n,1]]$ quantum stabilizer codes in this subsubsection.
Let $g=(a_0, a_1, \ldots, a_{n-1})$ be a vector of length $n$, odd or even,
with $a_0=0$ and $a_i=a_{n-i}=a_{-i\mod n}$ for $i=1$ to $\frac{n-1}{2}$.
We use $n\times n$ matrices $H_X, H_Z$ for convenience of explanation.
Let
\[H_X =\begin{bmatrix} 1 & 0 &  \ldots & 0& 1\\
                        0 & 1 &  \ldots &0 & 1 \\
                \vdots & \vdots   & \ddots & \vdots & \vdots \\
                        0 & 0 &  \ldots &1 & 1 \\
                        0 & 0 &  \ldots &0 & 0 \\\end{bmatrix}= \left[\begin{array}{cccc} &&&1\\&I_{n-1} &&\vdots \\ &&&1 \\0&\ldots &&0   \end{array}\right]
                \]
with ${(H_X)}_{i,i}={(H_X)}_{i,n-1} = 1$ for $0\leq i \leq n-2$ and ${(H_X)}_{i,j}= 0 $ otherwise.

Let ${H_Z}_{(i,j)}=a_{(j+1)\mod n}+a_{(i-j) \mod n}$ for $0\leq i, j \leq n-1$.
That is
\begin{align*}H_Z =&\begin{bmatrix} a_1+a_0 & a_2+a_{n-1} & a_3+a_{n-2}& \ldots & a_{n-1}+a_{2}& a_1\\
                       a_1+a_1 & a_2+a_{0} & a_3+a_{n-1}& \ldots & a_{n-1}+a_{3}&a_2\\
                       a_1+a_2 & a_2+a_{1} & a_3+a_{0} & \ldots & a_{n-1}+a_{4}&a_3\\
                       \vdots & \vdots & \vdots &\ldots & \vdots & \vdots\\
                        a_1+a_{n-2} & a_2+a_{n-3} & a_3+a_{n-4} & \ldots & a_{n-1}+a_{0}&a_{n-1}\\
                        a_1+a_{n-1} & a_2+a_{n-2} & a_3+a_{n-3} & \ldots & a_{n-1}+a_{1}&a_{0}
\end{bmatrix}\\
=&\begin{bmatrix} a_1 & a_2+a_{n-1} & a_3+a_{n-2}& \ldots & a_{n-1}+a_{2}& a_1\\
                       a_1+a_1 & a_2 & a_3+a_{n-1}& \ldots & a_{n-1}+a_{3}&a_2\\
                       a_1+a_2 & a_2+a_{1} & a_3 & \ldots & a_{n-1}+a_{4}&a_3\\
                       \vdots & \vdots & \vdots &\ldots & \vdots & \vdots\\
                        a_1+a_{n-2} & a_2+a_{n-3} & a_3+a_{n-4} & \ldots & a_{n-1}&a_{n-1}\\
                        0 & 0 & 0 & \ldots & 0 & 0
\end{bmatrix}.
\end{align*}
Then a check matrix $H=[H_X | H_Z]$ has rank $(n-1)$ and the commutative condition for $H$ can be justified as follows:
\begin{align*}
    \alpha_i\cdot \beta_j + \alpha_j\cdot \beta_i &=(\alpha_i+ \alpha_j)\cdot (\beta_i+\beta_j) \\
                                    &=(0\ldots 010\ldots 010\ldots 0 ) \cdot (\beta_i+\beta_j)\\
                                    &={H_Z}_{(i,i)}+{H_Z}_{(i,j)}+{H_Z}_{(j,i)}+{H_Z}_{(j,j)}\\
                                    &=a_{i+1}+(a_{i-j \mod n}+a_{j+1})+(a_{j-i \mod n}+a_{i+1})+ a_{j+1}\\
                                    &=0.
\end{align*}
We found that the quantum quadratic-residue code for $p=8j+5$ in the last subsubsection can be constructed in this way with $g$ being the indicator vector of the quadratic residues.
Some quantum codes achieving  the upper bound in \cite{CRSS98} can be constructed similarly.
For example, when $n=17$, each of the following two vectors
\[0110100110010110, \quad 0100011111100010 \]
(or their complementary vectors and no others) together with $a_0=0$ gives a $[[17,1,7]]$ code that achieves the upper bound in \cite{CRSS98}.
%
%
A $[[17,1,7]]$ quantum stabilizer code can be constructed by quantum BCH codes in \cite{GB99,AKS05}.
However, the above two vectors are found by computer search.
It is difficult to determine a vector $g$ and the minimum distance of the resulted quantum code efficiently.

\section{Conclusion}
    In this paper, a simple stabilizer code construction was proposed 
    based on syndrome assignment by classical parity-check matrices.
    The construction of quantum stabilizer codes can then be converted to the construction of classical binary linear block codes with commutative parity-check matrices.
    The asymptotic coding performance of this construction was shown to be 
    promisingly comparable to that of the CSS construction.

    Permutation matrices may help transform non-commutative parity-check matrices to commutative
    ones and/or increase the minimum distance of the constructed quantum codes.
    However, for a given parity-check matrix $H$, it remains open to find an effective permutation matrix $P$ such that $HP$ is commutative and/or corresponds to a code with greater minimum distance.

    We have constructed a family of stabilizer codes from classical binary Reed-Muller codes
    with performance comparable to that of the CSS construction.
    We have also investigated sufficient conditions for permutation matrices to be able to increase the minimum distance of our constructed quantum Reed-Muller codes by half.
    %
    %
    We have also proposed a specific kind of effective permutations and showed that
    they meet the sufficient
    conditions for stabilizer codes constructed from the $RM(1,m)$ Reed-Muller codes.
    A conjecture of effective permutation matrices for general $r,m$ with $m\ge 2r$ remains to be proved.
    It is believed that permutation of the columns of
    a parity-check matrix will play an important role in the construction of
    quantum stabilizer codes from classical parity-check matrices.


    The quantum quadratic-residue codes are codes with large quantum minimum distance.
    However their quantum minimum distance is hard to determine as in the classical case.
    Perhaps techniques in \cite{TCL05} can be applied to the quantum case.
    How to determine the minimum distance of a long quantum code will be a key to find good codes.

\bibliographystyle{IEEEtran}
\bibliography{IEEEabrv,qecc}

\vfill\eject
\end{document}